\documentclass{elsart}
\usepackage[T1]{fontenc}
\usepackage[latin1]{inputenc}
\usepackage{amsmath}
\usepackage{graphics}

\makeatletter

\providecommand{\LyX}{L\kern-.1667em\lower.25em\hbox{Y}\kern-.125emX\@}
\let\SF@@footnote\footnote
\def\footnote{\ifx\protect\@typeset@protect
    \expandafter\SF@@footnote
  \else
    \expandafter\SF@gobble@opt
  \fi
}
\expandafter\def\csname SF@gobble@opt \endcsname{\@ifnextchar[
  \SF@gobble@twobracket
  \@gobble
}
\edef\SF@gobble@opt{\noexpand\protect
  \expandafter\noexpand\csname SF@gobble@opt \endcsname}
\def\SF@gobble@twobracket[#1]#2{}

\usepackage{graphicx}

\usepackage[squaren]{SIunits}
\usepackage{color}
\usepackage{array}
\usepackage{dcolumn}
\usepackage{amssymb}
\usepackage{psfrag}
\usepackage{url}
\newcommand{\gr}[1]{\textcolor[gray]{0.6}{#1}}
\newcolumntype{d}{D{.}{.}{-1}}

\makeatother

\journal{Nuclear Instruments and Methods B}

\begin{document}

\begin{frontmatter}

\title{Measurement of the Enhanced Screening Effect of the d+d Reactions in Metals}

\author[tub]{A. Huke\corauthref{cor}}
\corauth[cor]{Institut für Optik und Atomare Physik,
Technische Universität Berlin, Sekr. PN 3-1, Hardenbergstr. 36, D-10623 Berlin, Germany}
\ead{huke@physik.tu-berlin.de, Armin.Huke@web.de}
\author[ustettin,tub]{K. Czerski}
\author[tub]{P. Heide}
\address[tub]{Institut für Optik und Atomare Physik,
Technische Universität Berlin, Berlin, Germany}
\address[ustettin]{Institute of Physics, University of Szczecin, Szczecin, Poland}

\begin{abstract}
The investigation of the d+d fusion reactions in metallic environments at sub-Coulomb
energies demands especially adapted techniques beyond standard procedures in
nuclear physics. The measurements which were performed with an electrostatic
accelerator at different self-implanted metallic target materials show an enhancement
of the reaction cross-section compared to the gas target experiments. The resulting
electron screening energy values are about one order of magnitude larger relative
to the gas target experiments and exceed significantly the theoretical predictions.
The measurements on deuterium inside metals are heavily affected by the interference
of two peculiarities of this system: the possibly very high mobility of deuterium
in solids and the formation of surface contamination layers under ion beam irradiation
in high vacuum systems. Thorough investigations of these processes show their
crucial influence on the interpretation of the experimental raw data. The differential
data acquisition and analysis method employed to it is outlined. Non observance
of these problems by using standard procedures results in fatal errors for the
extraction of the screening energies.
\end{abstract}


\end{frontmatter}

\section{Introduction}

A lasting topic in nuclear astrophysics is the screening of the nuclear Coulomb
potential due to the atomic electrons in laboratory nuclear physics experiments
which alterate the experimental reaction cross-sections at low energies relative
to the bare nucleus \cite{assenbaum87}. In contrast to the laboratory experiments
performed on gas targets, the screening effect in astrophysical plasmas is mainly
evoked by free electrons and leads in the case of strong coupling to an enhancement
of nuclear reaction rates in the interior of stars by many orders of magnitude
\cite{salpeter54,ichimaru93}. Indeed, the conditions of the stellar plasma
are hardly accessible for laboratory experiments. Nevertheless the knowledge
of the progression of nuclear reactions inside a plasma is crucial for the quantitative
understanding of stellar processes. We therefore investigated the d+d fusion
reactions in metallic environments where the electron gas can serve as a model
for a cold, dense, and strongly coupled plasma. As a result we first observed
a grossly enhanced screening effect wherein the screening energy is one order
of magnitude higher \cite{volos98,europhys01,dis} than in the case of a gaseous
deuterium target with \( \left( 25\pm 5\right) \, \electronvolt  \) \cite{greife95}.
Subsequent similar results were achieved by other authors \cite{yuki98,kasagi02,rolfs02,rolfs02b,rolfs03,rolfs04}.
The theoretical calculations \cite{europhys04} provide screening energy values
smaller by a factor of two compared to the experimental ones. On the other hand
these experimental data were achieved with standard data acquisition and analysis
methods \cite{rolfs88,rolfs02,kasagi02} and are inconsistent \cite{NPAII06b}.
The conditions in condensed matter are special because of the utmost high mobility
of the small hydrogen atoms. Therefore one can suppose neither a stable nor
a homogeneous deuteron density distribution which is required by the standard
analysis method. Furthermore the residual gas in the high vacuum systems common
in nuclear physics set-ups inevitably promotes ion beam induced chemical reactions
on the surface. Thus surface layers are built up consisting of metal oxide or
carbon deposits with different and varying deuteron densities. If this is not
taken into account fatal misinterpretation of the raw data is impending. To
address these problems we developed a specially adapted data analysis method
and thoroughly investigated the ion induced chemical surface reaction mechanisms
and their influence on the obtained screening energies which is outlined in
this work.

\section{\label{sec:experiment}Experiment}

\begin{figure*}[!htbp]
{\par\centering \resizebox*{1\textwidth}{!}{\includegraphics{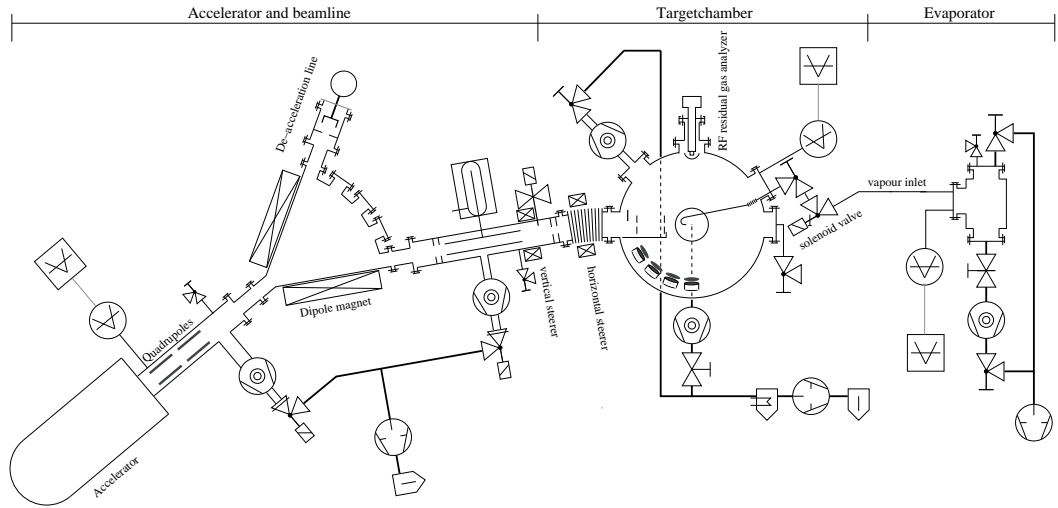}} \par}

\caption{\label{fig:aufbau}Experimental set-up}
\end{figure*}

An outline of the first experiment has already been given in Ref.~\cite{europhys01}
for which more details are given here and complemented by the modifications
and additions necessary for the in-depth investigation of the influences from
effects of surface physics. The set-up of the experiment is depicted in Fig.~\ref{fig:aufbau}
showing supplementary details of the vacuum system since this has special significance. 

The experiment has been performed at a cascade accelerator with a RF-ion source.
A highly stabilized \( 60\, \kilo \volt  \) power supply provided the acceleration
voltage. The high voltage corresponding to the beam energy was measured by a
precise voltage divider with an accuracy better than \( 1\, \volt  \). It was
mounted along the acceleration tube and includes the extraction voltage of the
ion source. The long term energy stability was about \( 10\, \electronvolt  \).
Since the cross-section is exponentially dependent on the projectile energy
it is particular important to control the energy determination. Therefore on
the opposite side of the beamline at the analyzing magnet an additional set-up
was installed with which the kinetic energy can be measured by means of the
opposite electric field method. A property of RF-ion sources is a voltage drop
in the plasma depending on the adjustment parameters which reduces the kinetic
energy of the ions below the extraction voltage \cite{kamke56}. Therefore the
beam energy is lower than inferred from the voltage divider by an average of
\( 120\, \electronvolt  \). Furthermore the ions leave the source with an energy
spread of \( 90\, \electronvolt  \) on an average due to the spatial extension
of the extraction zone \cite{dis}. This energy spread leads to an effective
positive energy shift in a thick target and for a nonresonant reaction \cite{europhys01,dis}.
However, it cannot compensate the energy loss from the voltage drop. The beam
consisting of \( \mathrm{D}^{+} \) and \( \mathrm{D}_{2}^{+} \)ions was focused
by a pair of electric quadrupoles and then analyzed in a magnetic dipole.

The beam line and the chambers of the vacuum system are made from aluminum with
elastomer rings at the joints. The vacuum is maintained by turbomolecular pumps
with auxiliary oil lubed two stage rotary vane pumps and can reach \( 3\cdot 10^{-7}\, \hecto \pascal  \)
at best in the target chamber and the cryogenic beamline. In such a high vacuum
system water vapour is the dominating constituent of the residual gas being
slowly desorbed from surfaces. A RF-quadrupole rest gas analyzer with a Faraday
cup detector was installed at the target chamber in order to monitor the composition
of the residual gas. The fraction of water can be distinctly reduced with cryogenic
traps. After the analyzing magnet the beam is led through a LN\( _{2} \)-cooled
Cu tube. The beam route inside the target chamber is covered by a LN\( _{2} \)-cooled
plate. Those cryogenic traps can remedy another problem provided the leakage
rate is not too high: It is known that oil from the rotary vane pumps can backstream
in low quantities even through modern turbomolecular pumps. This would lead
to a deposition of carbon in the beam spot.

\begin{figure}[!htbp]
{\par\centering \resizebox*{0.7\columnwidth}{!}{\includegraphics{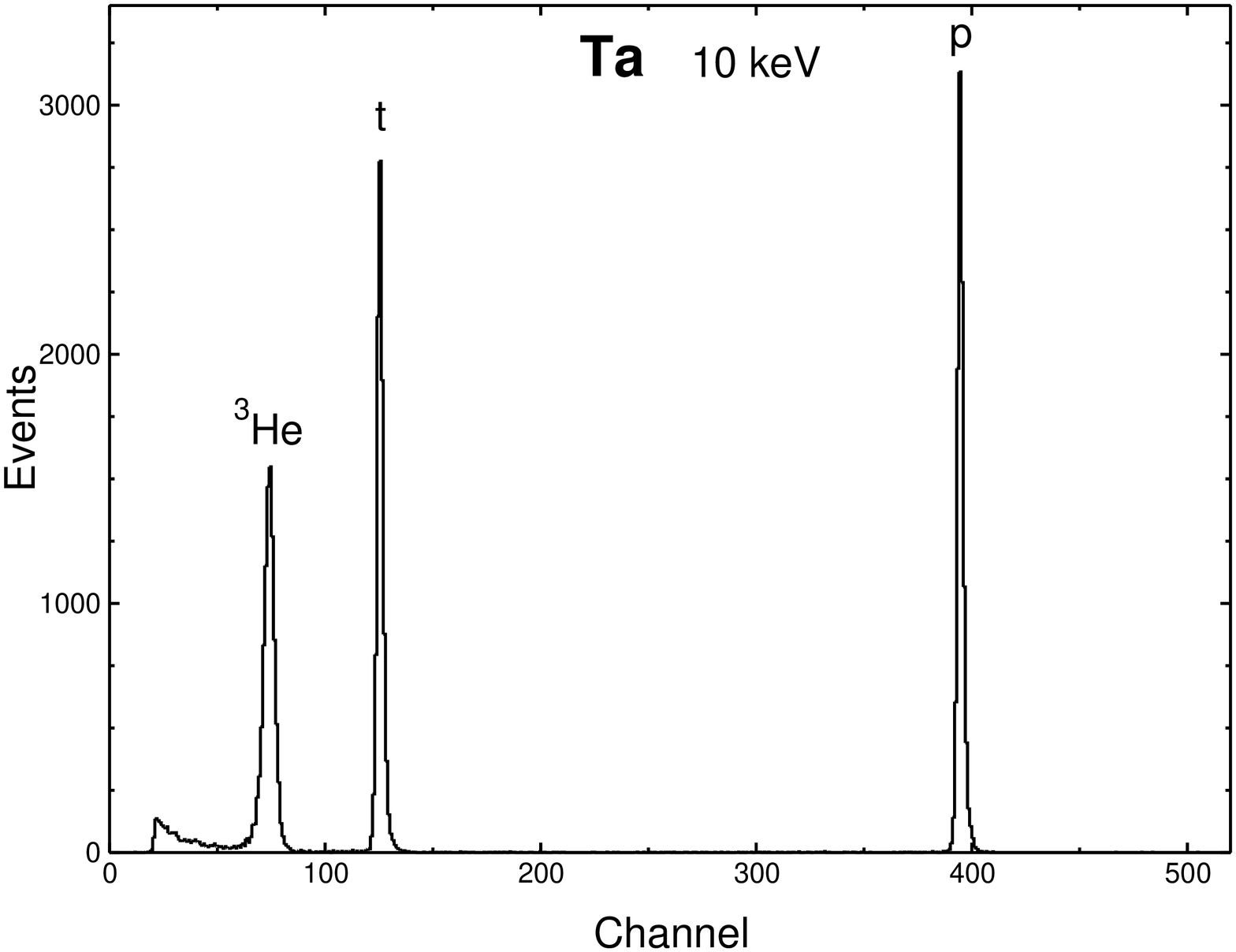}} \par}

\caption{\label{fig:ta10spectrum} Exemplary spectrum for \protect\( 10\, \kilo \electronvolt \protect \)
at Ta.}
\end{figure}

Directly in front of the target chamber the beam is deflected by an angle of
about 10\( ^{\circ } \) by a magnetic steerer in order to remove neutral particles.
In extension of the undeflected beam an isolated beam sink is mounted just behind
the entrance to the target chamber connected to an amperemeter. Therewith the
beam can be adjusted without disturbance of the deuteron density within the
target. The beam was focused on the target into a spot of about \( 1\, \centi \meter  \)
in diameter with the aid of apertures. The charge collected was determined via
the measurement of the electric current on the target holder which was isolated
from the target chamber. A negative voltage of \( 100\, \volt  \) was applied
to a surrounding metallic cylinder for suppression of secondary electrons. All
charged particles (\( ^{3} \)He, \( ^{3} \)H, p) from d+d reactions were detected
with four \( 100\, \milli \meter \squared  \) PIPS-detectors fixed at lab angles
of 90\( ^{\circ } \), 110\( ^{\circ } \), 130\( ^{\circ } \) and 150\( ^{\circ } \)
with respect to the beam in \( 10\, \centi \meter  \) distance from the target.
Aluminum foils of thickness \( 150\, \micro \gram \per \centi \meter \squared  \)
in front of the detectors prevented elastically scattered deuterons from entering
the detectors. Therefore the spectra allow for the clear distinction of all
three spectral lines as shown in Fig.~\ref{fig:ta10spectrum}. 

In order to be able to partially change the composition of the ambient gasmix
in a controlled manner (Sec.~\ref{sec:surfreactlayer}) a precise solenoid valve
is mounted at the target chamber which is used for the injection of gas. The
fraction of the infused gas is maintained at the desired level by feedback from
the vacuum gauge and the residual gas analyzer with a process computer. A detachable
conduit leads the gasflow to the target surface. Despite its openings the cylinder
around the target presents a significant obstacle for the molecular gas flow
in high vacuum. Therefore a turbomolecular pump is installed below the target
which exhausts this delimited volume. Non-gaseous agents can be transfered to
the gas phase in another vacuum chamber, the evaporator. However, certain restrictions
apply. The gas flow at the inlet of the solenoid valve must still be viscose
otherwise the conductivity approaches zero where no valve regulation is feasible
any more. Therefore the vapour pressure needs to be high enough. The other restriction
requires the vapour pressure of the agent to be higher than that of the pumping
oil of the rotary vane pump which is already fulfilled parenthetically by the
first restriction. The turbomolecular pump is used to exhaust the chamber for
cleanup before letting in the stuff to be gasified. The evaporator is a high
gas flow system which is pumped in normal operation at low vacuum by the rotary
vane pump only.

\section{\label{sec:anexpproc}Analysis and experimental procedure}

\subsection{Angular distribution}

A special property of the d+d reactions is the anisotropy of the angular distribution
down to the lowest energies which is due to the dominance of resonance levels
of negative parity in the \( ^{4}\mathrm{He} \)-system. Since the deuteron
has positive parity a spatial angular momentum of \( l=1 \) is required for
a coupling to these resonances that partially compensates the centrifugal barrier.
This anisotropy cannot be neglected even at the lowest deuteron energies. We
therefore used four detectors at different polar angles \( \theta  \) (in the
lab-system). Deuterons are bosons and appear as identical particles in the entrance
channel. Therefore the scattering amplitude must be symmetrized, which leaves
only Legendre polynomials with even \( l \) in the expression for the differential
cross-section. Therefore the angular distribution is symmetric to \( 90\degree  \).
After base transformation it is expressed by 
\begin{equation}
\label{eq:winkel.expans}
\frac{d\sigma }{d\omega }\left( \vartheta \right) =\sum _{k\in \left\{ \mathbb {N}_{0}\mid k\, \textrm{even}\right\} }a_{k}\cos ^{k}\vartheta \, ,
\end{equation}
 where \( \vartheta  \) is the scattering angle and \( d\omega  \) the solid
angle element in the CM-system. With four detectors only two addends can be
determined. The four counting numbers \( N\left( \theta \right)  \) gives with
the number of incident deuterons \( N_{0} \) an angle dependent yield \( Y \)
at the lab-energy \( E \) 
\begin{eqnarray}
\frac{1}{\varepsilon \Delta \Omega }\frac{dN(\theta )}{dN_{0}}=Y\left( E,\theta \right)  & \stackrel{!}{=} & n\frac{d\sigma }{d\omega }\left( \vartheta \right) \Delta x\label{eq:winkel.diffYdef} \\
 & = & n\frac{d\sigma }{d\omega }\left( \vartheta \left( \theta ,E\right) \right) \frac{d\omega }{d\Omega }\left( \theta ,E\right) \Delta x\, ,\nonumber \label{eq:winkel.diffYLab} 
\end{eqnarray}
 where \( \Delta \Omega  \) is the solid angle of the detector in the lab-system
and \( \varepsilon  \) its efficiency. On the right side the yield is given
by theory with the target particle density \( n \) and the target thickness
\( \Delta x \). Three problems arise from this expression: First, the differential
cross-section is given in the CM-System. In the second row the coordinate-transformation
is performed. \( \frac{d\omega }{d\Omega } \) is the transformation function
for the solid angle element between the lab-system (\( d\Omega  \)) and the
CM-system (\( d\omega  \)). Second, it is only valid for very thin targets.
For thick targets an integration over the depth \( x \) (and therewith the
projectile energy) in the target is required. Such would result in an expression
which is impracticable for a fit to the experimental data in order to determine
the \( a_{k} \). Third, in our case of autoimplanted targets the target particle
density is unknown. We therefore chose a procedure analog to Ref.~\cite{rolfs88}
where the coordinate transformation is performed at the beam energy. This is
justified by the conjuncture that the yield is dominated by reactions at the
highest energies, particularly in the sub-Coulomb energy range where the cross-sections
decreases exponentially. Therefore Eq.~(\ref{eq:winkel.diffYdef}) transforms
into 
\begin{equation}
\label{eq:winkel.Ntransf}
\frac{1}{\varepsilon N_{0}}\frac{1}{\frac{d\omega }{d\Omega }}\frac{N(\vartheta )}{\Delta \Omega }=n\Delta x\frac{d\sigma }{d\omega }\left( \vartheta \right) \, .
\end{equation}
 Introducing the differential counting number \( \frac{dN}{d\omega }(\vartheta ) \),
using Eq.~(\ref{eq:winkel.expans}) and resorting results in 
\begin{equation}
\label{eq:winkel.Ndiff}
\frac{dN}{d\omega }(\vartheta )=\varepsilon N_{0}n\Delta x\sum _{k\in \left\{ \mathbb {N}_{0}\mid k\, \textrm{even}\right\} }a_{k}\cos ^{k}\vartheta \, .
\end{equation}
 The product in front of the sum is constant for one measurement series and
can be included in the expansion coefficients \( a_{k} \). This model function
can now be fitted to the experimental data. The total counting number for the
full solid angle is then obtained by 
\begin{equation}
\label{eq:winkel.Ntotal}
N=\oint _{\mathbb {S}^{2}}\frac{dN}{d\omega }(\vartheta )\, d\omega =4\pi \sum _{k\in \left\{ \mathbb {N}_{0}\mid k\, \textrm{even}\right\} }\frac{1}{k+1}a_{k}\, .
\end{equation}
\begin{figure}[!htbp]
{\par\centering \resizebox*{1\columnwidth}{!}{\includegraphics{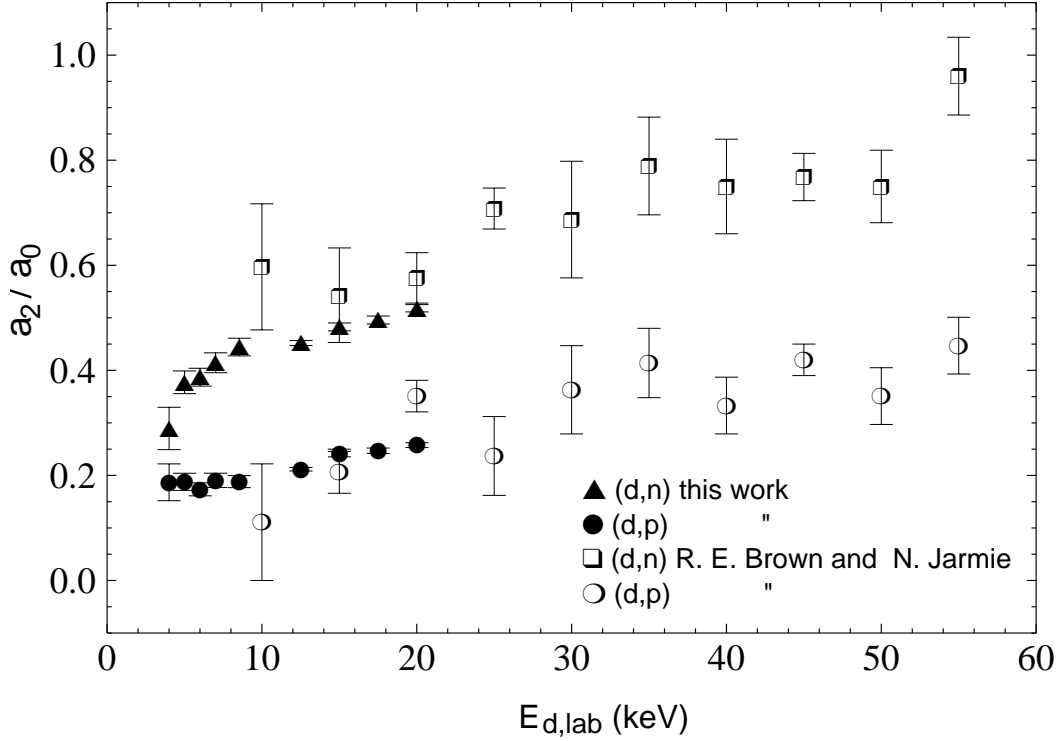}} \par}

\caption{\label{fig:aniso}Ratio of angular distribution coefficients \protect\( a_{2}/a_{0}\protect \)
in dependence of the deuteron energy in the laboratory for the two exit channels
in comparison to Ref.~\cite{brown90}.}
\end{figure}
 Fig.~\ref{fig:aniso} shows the ratio \( a_{2}/a_{0} \) for the case of tantalum
as host material in comparison to the gas target results from Ref.~\cite{brown90}.
They are in good agreement with the reference data, have noticeably smaller
errors and extend to lower energies. Therefore the previous statements conceive
confirmation.

\subsection{\label{sec:advanalysis}Differential analysis method}

An automatic (on-line) data acquisition system integrates the proton peaks of
the spectra at fixed time intervals (\( 15\, \second  \) - \( 2\, \minute  \))
and records the four counting numbers \( N\left( \theta \right)  \) and the
collected charge \( q \) in a file. With Eq.~(\ref{eq:winkel.Ntransf}, \ref{eq:winkel.Ndiff},
\ref{eq:winkel.Ntotal}) this can be reduced to the tabulated function \( N\left( q\right)  \).
The solid angle element \( \Delta \Omega  \) therein was determined with a
radioactive \( \alpha  \) calibration source (\( ^{241} \)Am). For the sake
of synopsis the data analysis method is depicted in a tabular form (\ref{eq:ytab}).
Quantities provided by the experiment are in column (\ref{eq:ytab}.A), terms
from theory in column (\ref{eq:ytab}.C). 

\begin{equation} 
\resizebox{\textwidth}{!}{
\definecolor{grau_dunkel}{gray}{0.4}
\definecolor{grau_hell}{gray}{0.8}
$\begin{array}{rcll@{\hspace{4em}}r}
\multicolumn{1}{c}{\text{\footnotesize A}} & \multicolumn{1}{c}{\text{\footnotesize B}} & \multicolumn{1}{c}{\text{\footnotesize C}} & \multicolumn{1}{c}{\text{\footnotesize D}} & \\
\multicolumn{1}{l}{\text{\scriptsize\sffamily\slshape\textcolor{grau_dunkel}{Experiment}}} & & \multicolumn{1}{c}{\text{\scriptsize\sffamily\slshape\textcolor{grau_dunkel}{Theory}}} & \multicolumn{1}{c}{\text{\scriptsize\sffamily\slshape\textcolor{grau_dunkel}{Supplements}}} & \\
\frac{1}{\varepsilon }\frac{dN}{dN_{0}}= & Y(E,q) & =\int\limits ^{R}_{0} n(q,x) \sigma (E(x))\, dx & & {\text{\footnotesize 1}} \\
\multicolumn{1}{l}{N_{0}=\frac{q}{ze}} & & \text{\color{black}\footnotesize \hspace{2em} substitution $E\rightarrow x$} & \textstyle \frac{dE}{dx}=-\left( \textcolor{grau_dunkel}{c_{M}}+\frac{\textcolor{black}{n(q)}}{n_{D}}\textcolor{grau_dunkel}{c_{D}}\right) \textcolor{black}{E^r} \: ,\; r\in \left\{ 0.45,\frac{1}{2}\right\} & {\text{\footnotesize 2}} \\
& & {\color{black} \text{\footnotesize \hspace{2em} $\sigma$ from \cite{brown90}} } & \textstyle \sigma (E)=2S_{0}\frac{1+\alpha E}{E}\textcolor{black}{\exp \left( -\frac{44.4021}{\sqrt{E}}\right)} & {\text{\footnotesize 3}} \\
\frac{ze}{\varepsilon }N^{\prime }(q)= & & = \colorbox{grau_hell}{$\frac{n(q)}{c_{M}+\frac{n(q)}{n_{D}}c_{D}}$} \fcolorbox{grau_dunkel}{white}{$\int\limits ^{E}_{0}\frac{\sigma (E)}{E^{r}}dE$} & & {\text{\footnotesize 4}} \\
& {\Huge \Downarrow} & \hspace{3em} \textcolor{grau_dunkel}{\hookrightarrow n(q)} & & \\
\multicolumn{2}{r}{ \fbox{$\frac{Y(E;q)}{\int\limits ^{E}_{0}\frac{\sigma (E)}{E^{r}}dE}=:y(E;q)$} } & = \textcolor{grau_dunkel}{\frac{n(q)}{c_{M}+\frac{n(q)}{n_{D}}c_{D}}}\times \textcolor{black}{F(E)} & & {\text{\footnotesize 5}} \\
& {\Huge \Updownarrow} & & & \\
\frac{y_{\mathrm{mod}}(E)}{y(E)}= & F(E) & =\frac{Y_{\mathrm{mod}}(E)}{Y(E)} & \multicolumn{1}{l}{Y_{\mathrm{scr}}(E)=\frac{n(q)}{c_{M}+\frac{n(q)}{n_{D}}c_{D}}\int\limits ^{E}_{0}\frac{\sigma (E+2\textcolor{black}{U_{e}})}{E^{r}}dE} & {\text{\footnotesize 6}} \\
& {\Huge \Downarrow} & & & \\
\multicolumn{2}{r}{\fbox{$\frac{y(E)}{y(\textcolor{grau_dunkel}{E_{0}})}=:F_{\mathrm{norm}}(E)$}} & =\frac{F(E)}{F(\textcolor{grau_dunkel}{E_{0}})} & & {\text{\footnotesize 7}} \\
& {\Huge \Downarrow} & & & \\
& {\Huge \textcolor{black}{U_e}} & & & \\
\end{array}$
} \label{eq:ytab}
\end{equation}

The total yield is now evaluated according to (\ref{eq:ytab}.1). Starting from
\( N\left( q\right)  \) the yield calculates as (\ref{eq:ytab}.A1) with the
efficiency \( \varepsilon  \) of the detector which is 1 for PIPS-detectors.
The number of incident deuterons \( N_{0} \) is expressed by the collected
charge \( q \) with (\ref{eq:ytab}.A2) where \( z \) is the charge state
of the projectile which is 1 for \( \mathrm{D}^{+} \) and \( \frac{1}{2} \)
for \( \mathrm{D}_{2}^{+} \). This leads to (\ref{eq:ytab}.A4). Due to the
high mobility of hydrogen in metals which is many orders of magnitude greater
than for larger atoms \cite{mhydrides68} the stability of the target particle
density is not guaranteed any more. Therefore the usual assumption of a uniform
and time invariant particle density is not applicable for this situation. Consequently
the derivative \( \frac{dN}{dN_{0}} \) (\ref{eq:ytab}.A1) must not be simply
replaced by the difference quotient \( \frac{\Delta N}{\Delta N_{0}}=Y \) for
the yield of one measurement in the usual manner. Instead a changing yield function
\( Y(q)\sim N'\left( q\right)  \) (\ref{eq:ytab}.AB4) needs to be retained.
Otherwise misinterpretation of the raw data is impending (Sec.~\ref{sec:discuss}).
The applied discretization increment time equivalent of the function \( N\left( q\right)  \)
is short in comparison to the implantation and diffusion times. The deuteron
density in (\ref{eq:ytab}.C1) is likewise dependent on \( q \) and in general
on the target depth \( x \). The targets are infinitely thick in relation to
the range \( R \) of the projectile ions. Therefore the integration over the
depth is required.

The depth \( x \) on its part can be substituted by the ion energy using the
linear stopping power relation (\ref{eq:ytab}.D2). The term in brackets contains
the stopping power coefficients for the host metal \( c_{M} \) and deuterium
\( c_{D} \) whereby the later is weighted by the ratio of the deuterium density
in the target to the density of liquid deuterium. The stopping power function
\( E^{r} \) is valid for the energy range below \( \beta \simeq 0.01 \) where
the Lindhard theory \cite{lindhard54,lindhard61,lindhard63} provides a good
explanation with \( r=\frac{1}{2} \). The numerical value of the stopping power
coefficients can be taken from the compilation of Ref.~\cite{ziegler77}. They
also state that an exponent of \( r=0.45 \) fits the experimental data better.
For our purposes this difference has low impact. The stopping power coefficients
have an error between \( 10-20\, \% \). It is therefore highly desirable to
manage without them. According to Braggs rule \cite{bragg05} only the mass-specific
stopping powers can be linearly added. In this special case however where the
hydrogen number density is set in relation to the metal number density (i.e.~\( \mathrm{MD}_{\mathrm{x}} \))
a brief calculation shows that (\ref{eq:ytab}.D2) holds. Since \( n \) is
unknown in (\ref{eq:ytab}.C1), a reference for the cross-section \( \sigma  \)
is required which is provided by Ref.~\cite{brown90} as essentially a S-factor
parameterization (\ref{eq:ytab}.D3). Their measurements have with an error
of \( 1.5\% \) by far the highest precision. In this case it is presupposed
that the density \( n \) is independent of the depth \( x \) a factorized
expression for \( Y \) accrues (\ref{eq:ytab}.C4) with a charge dependent
term (gray shaded) and the energy integral (framed). The charge dependent term
contains the density \( n\left( q\right)  \) which can be calculated using
the stopping power coefficients. Before the measurement we implanted our targets
up to a saturation level in the proximity of the stoichiometric ratios known
from physical chemistry \cite{mhydrides68} (\( \log _{10}n\cong 22-23 \)).
Under these conditions the homogeneity of the density is substantiated. We will
reinvestigate this assumption in Sec.~\ref{sec:model}.

The derivation up to now serves as a motivation for the introduction and definition
of the \emph{reduced yield} \( y \) (\ref{eq:ytab}.AB5) which originates from
the previous equation (\ref{eq:ytab}.4) by division with the energy integral.
The experimental yield in (\ref{eq:ytab}.A5) is calculated from (\ref{eq:ytab}.A4).
A numerical differentiation is considered instable which is additionally aggravated
because the counting number \( N\left( q\right)  \) is scattered by the counting
statistics\footnote{%
Markoff's formula is not applicable under this circumstances \cite[chapter 25]{abramowitz}.
}. Hence we used cubic spline polynomials with natural border conditions in order
to calculate the derivative at the \( q_{i} \)'s from their monomial coefficients.
This equalization function is well defined also at the borders of the interval.
The charge dependent term containig the density \( n\left( q\right)  \) in
(\ref{eq:ytab}.C5) indicates that \( y \) is constant for all energies and
independent of the implanted charge if the previous assumption of a homogeneity
is valid and the experimental yield does not deviate from the functional progression
in the energy integral (\ref{eq:ytab}.A5). Therefore the reduced yield is a
sensible indicator for any such deviation, be it changes in the number density,
the stopping power or effects which modify the cross-section like screening.
Which of these is efficacious is a matter of interpretation and is determined
by the theoretical expression for the yield. Since our observations showed a
strong energy dependence of \( y \), we take this effect into account by the
introduction of an energy dependent factor \( F(E) \) in (\ref{eq:ytab}.C5).
Obviously \( F \) can be calculated by the quotient of a reduced yield \( y_{\mathrm{mod}} \)
modified in the said manner and the unmodified reduced yield (\ref{eq:ytab}.A6).
This is equal to the quotient of the corresponding yields (\ref{eq:ytab}.C6)
since the energy integrals from (\ref{eq:ytab}.A5) shorten themselves. If the
modification is attributed to electron screening the screened yield can be written
as (\ref{eq:ytab}.D6) in close analogy to Ref.~\cite{assenbaum87} where the
screening energy \( U_{e} \) is added to the kinetic energy. The factor 2 arises
from the CM-Lab-transformation. This would result in 
\begin{equation}
\label{eq:Fthick}
F(E)=\frac{\int\limits ^{E}_{0}\frac{\sigma (E+2U_{e})}{E^{r}}dE}{\int\limits ^{E}_{0}\frac{\sigma (E)}{E^{r}}dE}
\end{equation}
 which shows that \( F \) is an enhancement factor for thick targets in analogy
to the enhancement factor for thin targets from Ref.~\cite{assenbaum87} 
\begin{eqnarray}
f(E_{\mathrm{CM}}) & := & \frac{\sigma \left( E_{\mathrm{CM}}+U_{e}\right) }{\sigma \left( E_{\mathrm{CM}}\right) }\label{m:kontext.enhancedef-f} \\
 & = & \frac{\frac{1}{E_{\mathrm{CM}}+U_{e}}S\left( E_{\mathrm{CM}}+U_{e}\right) e^{-2\pi \eta \left( E_{\mathrm{CM}}+U_{e}\right) }}{\frac{1}{E_{\mathrm{CM}}}S(E_{\mathrm{CM}})e^{-2\pi \eta (E_{\mathrm{CM}})}}\nonumber \\
 & \simeq  & e^{\left( \pi \eta \left( E_{\mathrm{CM}}\right) \frac{U_{e}}{E_{\mathrm{CM}}}\right) }\qquad ,\, U_{e}\ll E_{\mathrm{CM}}\: ,\nonumber 
\end{eqnarray}
 using the S-factor parameterization of the cross-section in the second line
and applying an approximation in the third line, which demonstrates its qualitative
behaviour as a roughly exponential increase for decreasing energies.

\begin{figure}[!htbp]
{\par\centering \resizebox*{1\columnwidth}{!}{\includegraphics{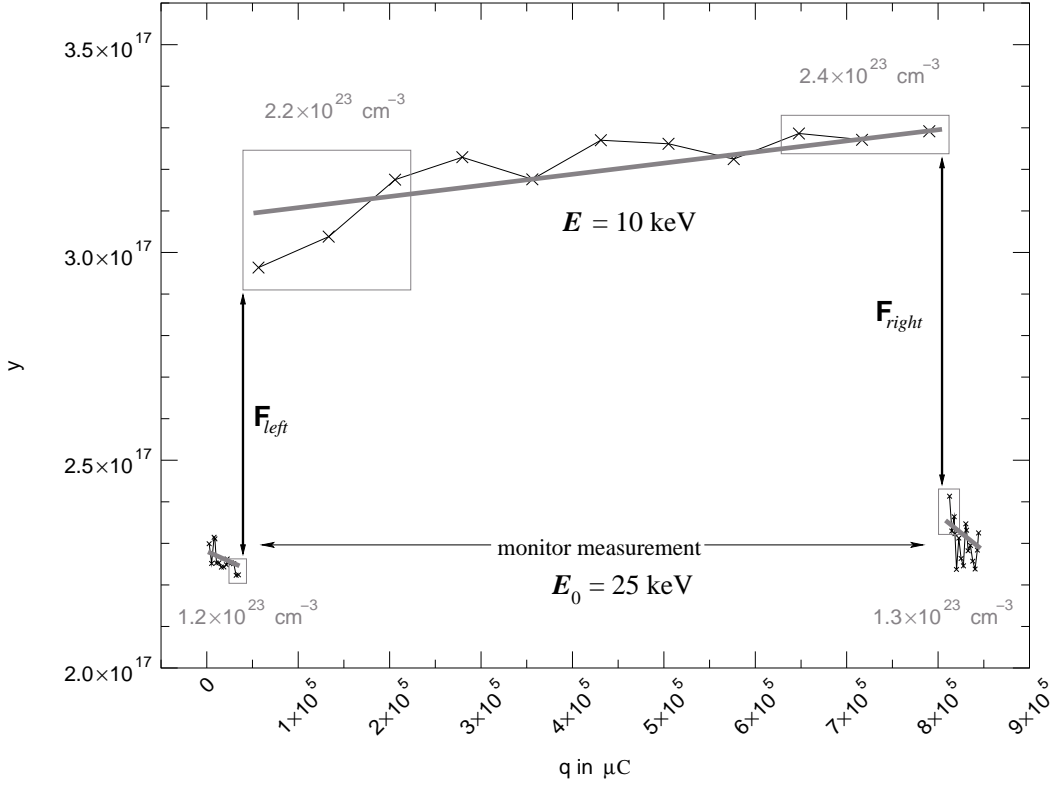}} \par}

\caption{\label{fig:expprocedure}Analysis procedure at the example of Zirconium at
\protect\( 10\, \kilo \electronvolt \protect \)}
\end{figure}

Because the unmodified reduced yield is unknown we use a normalization at a
fixed energy \( E_{0} \) what is implemented by the definition of the \emph{normalized
enhancement factor} \( F_{\mathrm{norm}} \) in (\ref{eq:ytab}.AB7). This definition
is equally free from restrictive presuppositions as the one for the reduced
yield (\ref{eq:ytab}.AB5). Obviously (\ref{eq:ytab}.C7) follows from (\ref{eq:ytab}.C5).
Fig.~\ref{fig:expprocedure} demonstrates the practical analysis procedure. 
 We chose the normalization energy at \( E_{0}=25\, \kilo \electronvolt  \)
because the effects of the screening enhancement are low here. The reduced yields
of a measurement at the energy \( E \) are consecutively compared to these
of the monitor measurements at \( E_{0} \). The straight lines are just for
guidance. The normalized enhancement factor is determined by the ratio of the
reduced yields at the discontinuities (Fig.~\ref{fig:expprocedure} \( F_{left} \)
and \( F_{right} \)). Since the measurements used for the data pooling are
taken at the saturation density close to the stoichiometric ratio and the switch
of the projectile energy followed by the measurement of the reduced yield happened
very quickly (\( \lesssim 4\, \minute  \)) the discontinuity cannot be caused
by a change in the target deuteron density perhaps due to a different projectile
flux or energy deposition in the target. The gray rectangles indicate the points
from which the error for \( F_{\mathrm{norm}} \) is inferred. Thus not only
errors from the counting statistics are included but also from other deviations
of the reduced yield e.g.~long term changes of the density. Therefore from line
(\ref{eq:ytab}.6) on the charge dependency drops out. We attribute the discontinuity
to the screening effect whereas the long term changes of \( y\left( E;q\right)  \)
at a fixed energy \( E \) are caused by the development of the density \( n\left( q\right)  \).
In that way we can discern both causes for the variations of \( y \) which
is not possible if only total yields were compared. We also do not need to know
the absolute value of the particle density \( n \) and are independent of the
stopping power coefficients. Accordingly the reduced yield is described with
Eq.~(\ref{eq:ytab}.D6) and results in the theoretical expression for the normalized
enhancement factor 
\begin{equation}
\label{m:ered.FnormTheo}
F_{\mathrm{norm}}(E)=\frac{\int\limits ^{E}_{0}\frac{\sigma (E+2U_{e})}{E^{r}}dE}{\int\limits ^{E_{0}}_{0}\frac{\sigma (E+2U_{e})}{E^{r}}dE}\frac{\int\limits ^{E_{0}}_{0}\frac{\sigma (E)}{E^{r}}dE}{\int\limits ^{E}_{0}\frac{\sigma (E)}{E^{r}}dE}
\end{equation}
 with the single fit parameter \( U_{e} \). The theoretical curve for \( F_{\mathrm{norm}} \)
is in good agreement with the experimental points as Fig.~\ref{fig:fituxim}
(right, bottom) shows and consequently confirms the implied assumptions. So
far \( U_{e} \) is used as an energy shift parameter which can be a sum of
several effects not exclusively screening. The reduced yield can be used to
calculate a density estimate by solving Eq.~(\ref{eq:ytab}.BC5) towards \( n\left( q\right)  \)
with the inherent suppositions and ignoring \( F \):
\begin{equation}
\label{m:red.nDichtey}
n(q)=\frac{c_{M}n_{D}y(E;q)}{n_{D}-c_{D}y(E;q)}.
\end{equation}
 The gray numbers in Fig.~\ref{fig:expprocedure} represent average densities
for the rectangles below computed in such a way. The stopping power coefficients
are exclusively necessary for it. Because of the nonlinear relation between
\( y \) and \( n \) the magnitude of the discontinuities between the energies
is different, too.

\section{\label{sec:surfreactlayer}Beam induced surface reactions and layer formation}

Apart from other possible error sources, particularly the energy definition,
it turned out that the dominating error source is the formation of layers at
the target surface. Therefore this subject needs special experimental and theoretical
treatment in order to recognize its repercussions and realize the underlaying
mechanisms.

\subsection{\label{sec:surfeff}Surface effects}

The processes taking place at the target can be understood in terms of vacuum
and surface physics, chemistry and nuclear radiotomy.

It is well known in nuclear physics that oil from the forepumps is cracked up
and deposited as carbon layers inside the beam spot onto the target surface.
This nuisance can easily be avoided by the use of LN\( _{2} \)-cooled cryogenic
traps. But there is another problem that has usually no attention: Water molecules
adsorbed at the target surface can be dissociated under the impacting beam ions
and the produced oxygen radicals in chemisorption form strongly bonded metal
oxide layers at the surface. This process is progressing into the bulk of the
target material along the damage track of the de-accelerating ions. It is even
speedup by an effect known as \emph{embrittlement} from the physical chemistry
of the metal hydrides \cite{mhydrides68} which means that the crystal structure
of the metal is bursted by the recrystallization process that accompanies the
formation of the metal hydride crystal thus preparing the way for the oxygen
radicals. Therefore the metal oxides can build up to a thickness of more than
100 atomic layers. Since oxygen has a higher affinity to metals than hydrogen
no metal hydride formation happens any more. Instead hydrogen is segregated
in the metal oxide with very low and instable densities. Since the hydrogen
depleted zone has a higher impact for lower beam energies on the integral in
Eq.~(\ref{m:ered.FnormTheo}) the normalized enhancement factor \( F_{\mathrm{norm}} \)
becomes \( <1 \) for \( E<E_{0} \) (Sec.~\ref{sec:model}). Finally no screening
is visible under this circumstances. When measuring reactions on metallic nuclei
metal oxide layers at the surface are not an issue because the beam energies
are in the high \kilo\electronvolt and \mega\electronvolt region where the range
of the ions in the target is high in comparison to the thickness of the layer.
Moreover the oxidation only loosens the surface structure \footnote{%
Very light metals like lithium and beryllium are used in form of their salts
that cannot oxidize of course.
}. Therefore the situation in metal hydrides is singular where very low energies
are required for the investigation of the screening at deuterium in combination
with the high stopping powers of the metals and that is why contamination layers
are fatal (Sec.~\ref{sec:model}).

\begin{figure}[!htbp]
{\par\centering \resizebox*{1\columnwidth}{!}{\includegraphics{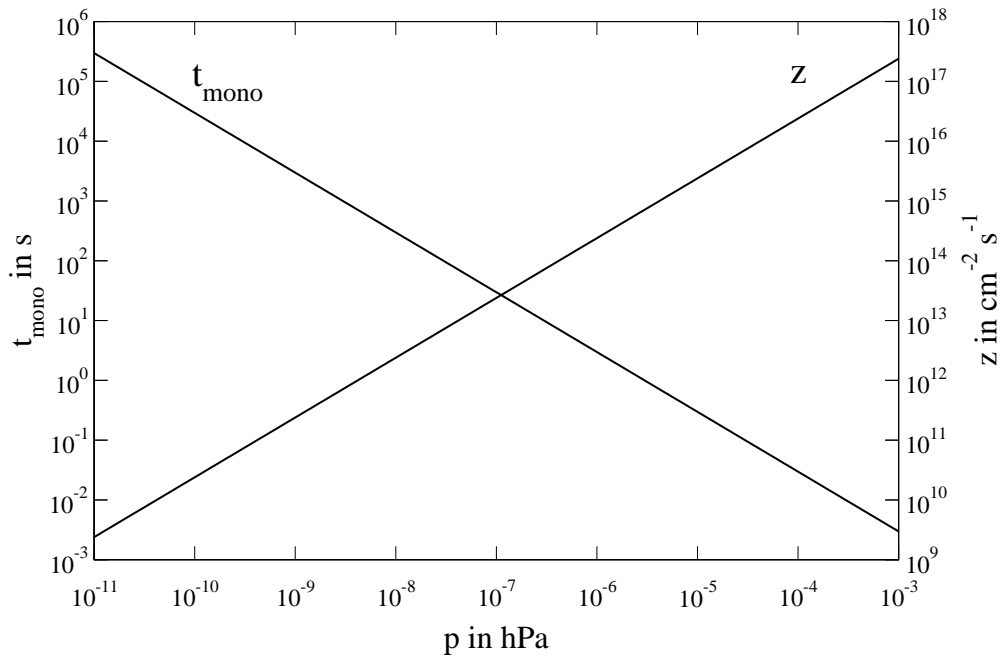}} \par}

\caption{\label{fig:H2Ovapor}Gaskinetic quantities of water vapor}
\end{figure}

Vacuum components that has been exposed to room ambient for extended times may
contain up to 100 monolayers of water vapour. The desorption rate of water at
room temperature is so low that it cannot be pumped off from a high vacuum system
in practicable periods of time without baking but remains as the dominating
constituent of the residual gas. Fig.~\ref{fig:H2Ovapor} illustrates the situation
at the target. 
 Of interest are the hit rate \( z \) at \( 300\, \kelvin  \) \footnote{%
\( z=\frac{1}{6}p\sqrt{\frac{8}{\pi kTM_{\mathrm{water}}\atomicmass }} \) with
the molecular weight of water in atomic mass units \atomicmass.
} and the time \( t_{\mathrm{mono}} \) it takes to build up one monolayer of
water molecules at the target surface \footnote{%
\( t_{\mathrm{mono}}=\left( zSd^{2}\right) ^{-1} \) with the distance between
the water molecules at the target surface assumed as \( d=3.75\cdot 10^{-10}\, \meter  \)
and the sticking coefficient \( S=1 \). 
}. In the pressure interval \( 10^{-8}-10^{-6}\, \hecto \pascal  \) the hit
rate is in the order of magnitude \( 10^{12}-10^{14}\, \centi \meter \rpsquared \reciprocal \second  \).
This is comparable to the ion flux of \( \simeq 4\cdot 10^{13}\, \centi \meter \rpsquared \reciprocal \second  \)
for a current of \( 10\, \micro \ampere  \) with a beam diameter of \( 1\, \centi \meter  \).
Therefore there is sufficient supply for the oxidation process even if the target
was originally clean. The time for mono layer formation is in the range \( 2-200\, \second  \)
for the above pressures. Withal is assumed that water has a sticking coefficient
of 1.

Due to its extraordinarily high dipole momentum of \( 1.84\, \textrm{Debye} \)
water has a high binding energy to solids and is hence chemisorbed to surfaces
with preferable the oxygen atom oriented towards the surface. There are also
counteracting processes like thermal or ion stimulated desorption and sputtering.
Heating the material is the way of choice in vacuum technologie to speed up
the desorption of water from the surfaces. This works well for the upper layers
but for water molecules at the joint face to the solid there is another option
aside from being desorbed \cite{henzler91,zangwill88}: The chemisorbed water
molecule can simplified be described by a Lennard-Jones type potential. There
is another such potential which describes a chemisorbed oxygen atom with its
minimum closer to the surface and deeper. The intersection of this two potentials
defines the activation energy barrier \( E_{a} \) to dissociative chemisorption,
i.e.~the protons are splitted off so that the oxygen atom can form a firmer
bond to the metal \cite{chabal84,scheffler84}, what eventually leads to metal
oxide formation. An interesting and apparently paradox phenomenon is now that
the probability \( \sigma  \) for sticking and dissociative chemisorption becomes
greater with raising temperature if \( E_{a}>0 \) \cite{zangwill88}. In this
case heating proliferates the oxidation process which is only stopped by the
condition that the change of the free enthalpy after and prior to the surface
reaction \( \Delta G=\Delta H-T\Delta S \) must be negative. Since the entropy
difference \( \Delta S \) is negative for the adsorption and positive for desorption
the process will eventually turn over. If \( E_{a}<0 \) desorption is preferred.
The impacting beam heats the target considerably because the beam energy is
dissipated in a volume with a thickness of \( \lesssim 1\, \micro \meter  \).

But besides heating and phonon generation the traversing ions can directly interact
with the adsorption complex by excitation and ionization providing activation
energy and thereby enabling chemical reactions which under normal conditions
will not occur. Therefore even noble metals can be oxidized this way. If a projectile
hit removes an electron from a binding orbital the adsorbed particle may suddenly
be via a Franck-Condon transition on the repulsive part of an ionic potential.
It is accelerated down the slope and arrives as a positive ion at the gas phase.
This would be ion stimulated desorption. However the ionic potential curve has
intersections with a manifold of excited states of the adsorption complex where
the particle can jump to any one of these which can eventually lead to dissociative
chemisorption again. In either case it depends on the concrete form of the interaction
potential whether dissociative chemisorption or desorption prevails. This means
that there is a strong material dependency but the ion energy and flux exert
a significant influence too. Generally reactive metals have a higher oxidation
rate then less reactive ones. As a matter of principle sputtering can clean
surfaces by removal of the top atomic layers. This is here a side effect of
the irradiation. We could observe on surfaces in the viewing solid angle of
the beam spot blue brown coloured films. An EDX-analysis showed that those films
consist of oxygen. Therefore sputtered oxygen radicals produced metal oxides
on other surfaces. Yet the sputtering yield of the lightweight deuterons is
too low to keep the metal surface clean; the ion stimulated oxidation prevails.

So far the oxidation of the metal with the vanishing of the screening effect
may seem inevitable in high vacuum. Fortunately there is a process that can
prevent the oxidation. Carbon hydride molecules can be physisorbed at the surface,
dehydrated under ion irradiation and the remaining carbon radicals can react
with oxygen to carbon monoxide which leaves the surface into the gas phase.
The metal surface acts as a catalyzator that condenses and concentrates the
two agents for the ion induced chemical reaction and remains unchanged. The
difficulty in using this mechanism in order to keep the target clean is that
this works only in a narrow and labile equilibrium for the dosages of water
and the carbonhydride. Owing to the tetrahedral structure of the carbon sp\( ^{3} \)-hybrid
orbitals carbonhydrides are unipolar. Consequently, they can interact through
van-der-Waals forces and be physisorbed to surfaces only. Ion irradiation causes
a rupture of the C-H-bond in carbonhydrides \cite{hering99}. The residual carbonhydride
radicals with unsaturated C-bonds are very reactive and are now chemisorbed
at the surface where the dehydration by ion impact can be continued. For this
carbon adatom generation with subsequent CO-reaction to take place the initial
break up of the C-H-bond must happen in the physisorbed state. Physisorption
is usually a weak bond. Thermal and beam stimulated desorption antagonize in
contrast to the situation of water. Regarding the n-alcanes every CH\( _{2,3} \)-group
adds a bonding energy of \( 40-60\, \milli \electronvolt  \) through van-der-Waals
interaction. With increasing length of the chain the intermolecular interaction
can achieve considerable magnitudes and becomes even greater than the intramolecular
forces at the C-C-bond. Large n-alcanes do not perform phase transitions any
more but disintegrate instead. Therefore the phase transition temperatures (or
just as well the vapour pressures) are a measure for the strength of the total
van-der-Waals force of the molecule and accordingly for the physisorption what
can as well be illustrated by the mirror charges induced into the metal surface.
For longer alcanes the process will work better than for shorter ones.

\subsection{\label{sec:gasadmix}Gas admixture experiments}

A very expandable experimental campaign provided evidence for the previous picture
of the processes at the target which explains the encountered phenomena.

As already said not using the cryogenic traps causes a gray carbon film to be
deposited in the beam spot. The carbon film is implanted with deuterium, too
whereby densities of \( 3-5\cdot 10^{22}\, \centi \meter \rpcubed  \) can be
achieved. Interestingly enough carbon does not show any discontinuities of the
reduced yields which indicate screening. Two separate measurement series with
different carbon film preparation techniques verified that. The pumping oil
is so effective that small traces are enough in order to build a carbon layer.
The oil fragments were hardly above the detection threshold of the RGA. The
pumping oil is Balzers~P3 and has a molecular mass of \( \simeq 500\, \atomicmass  \)
which would correspondend to 35 carbon atoms and a vapour pressure of \( 7\cdot 10^{-5}\, \hecto \pascal  \)
at \( 20\celsius  \) \footnote{%
Balzers did not want to provide information about the chemical structure.
}. The computer regulated solenoid valve was used to inject oxygen (O\( _{2} \))
into the target chamber while irradiating an aluminium target in order to compensate
the oil excess. Even at the highest pressure that was justifiable with this
pump arrangement no detectable change in the implantation behaviour could be
observed. Therefore gaseous molecular oxygen had no effect on the layer formation.
Using the evaporator to inject water vapour a moderate increase of the pressure
was sufficient in order to oxidize the target surface. On the other hand with
the active cryogenic trap methane (CH\( _{4} \)) was given a trial. Methane
had likewise no effect. Only at the highest pressure of \( 5\cdot 10^{-5}\, \hecto \pascal  \)
some carbon precipitated at the rim of the beam spot where the temperature and
as a result desorption is lower. The results are in good agreement with the
above statements. Unipolar weakly bound molecules like oxygen and methane are
preponderantly desorbed by the beam and do not contribute to the surface reactions
whereas water proofed to be the source for the oxidation of the metals. The
viscose pumping oil is highly effective in providing carbon radicals at the
surface. Nonetheless it is not easy to achieve the desired equilibrium with
it mainly because it is effective close to the detector threshold and it cannot
be handled with the injector. For this purpose a RGA with electron multiplier
detector and a high vacuum dosing valve would be required. Both were not available.
Besides the partial pressure of the oil was not stable so operating with a regulated
water vapour injection alone had no lasting benefit, too.

The alkane with the lowest vapour pressure that fits into the design restrictions
of the evaporator including a safety margin is decane (C\( _{10} \)H\( _{22} \)).
Its vapour pressure is \( 1.39\, \hecto \pascal  \) at \( 20\celsius  \).
Putting up with possible penalties due to its lower interaction force in comparison
to oil it was the best compromise with the given equipment. With varying decane
partial pressure the equilibrium where the target surface keeps clean was then
framed. 
\begin{figure}[!htbp]
\resizebox*{1\columnwidth}{!}{\includegraphics{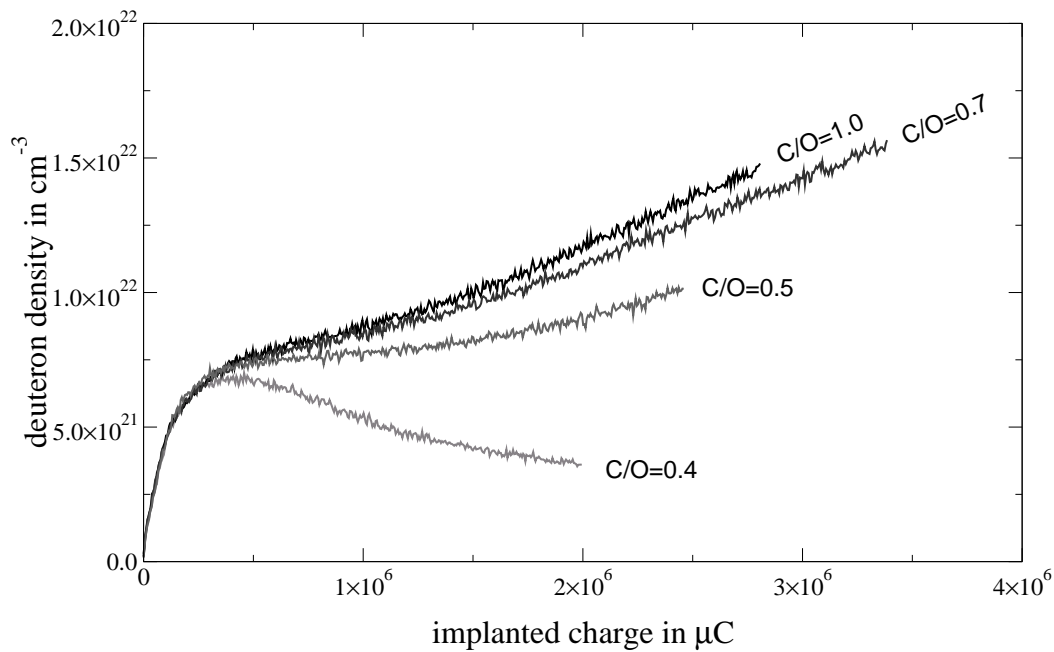}}

\caption{\label{fig:AlImplant}Influence of the composition of the residual gas on the
implantation for aluminum. The curves are for different C/O-ratios produced
by different mixtures of decan and water.}
\end{figure}
 Fig.~\ref{fig:AlImplant} shows the initial implantation curves for different
C/O-ratios on aluminium with a beam current of \( 30\, \micro \ampere  \).
The C/O-ratio is defined by the partial pressures of decane and water as \( \mathrm{C}/\mathrm{O}=10P(\mathrm{C}_{10}\mathrm{H}_{22})/P(\mathrm{H}_{2}\mathrm{O}) \).
Since the partial pressures are measured by the RGA far away from the target
with significant obstacles for the molecular flow in between the actual value
of the C/O-ratio at the target can deviate considerably. The deuteron densities
are calculated with Eq.~(\ref{m:red.nDichtey}) from the tabulated reduced yield
function \( y(q) \) provided by the automatic data acquisition system. Eq.~(\ref{m:red.nDichtey})
implies the supposition of a homogenous depth distribution what is not valid
for the initial implantation process before saturation of course. So for the
non-saturated case the density from Eq.~(\ref{m:red.nDichtey}) is rather to
be interpreted as a depth averaged value. The lowest curve represents the behaviour
during the creation of a metal oxide layer. The deuteron density remains an
order of magnitude below the stoichiometric ratio. It reaches a maximum at the
beginning for more reactive metals like aluminium in this example. Increasing
C/O-ratios lead also to an accretion of the densities which continue up to the
saturation density at the stoichiometric ratio. But too high C/O-ratios causes
the build up of a carbon layer covering the metal. Here the ratio \( 0.5 \)
is within the equilibrium window.

The C/O-ratio is the command variable in a cybernetic cascade regulating circle
implemented on a process computer with the RGA as receptor. It is thereby held
stable. The manipulated variable is the total pressure which is in turn the
command variable for the inner regulating circle with the solenoid valve as
effector. 
\begin{figure}[!htbp]
{\par\centering \resizebox*{1\columnwidth}{!}{\includegraphics{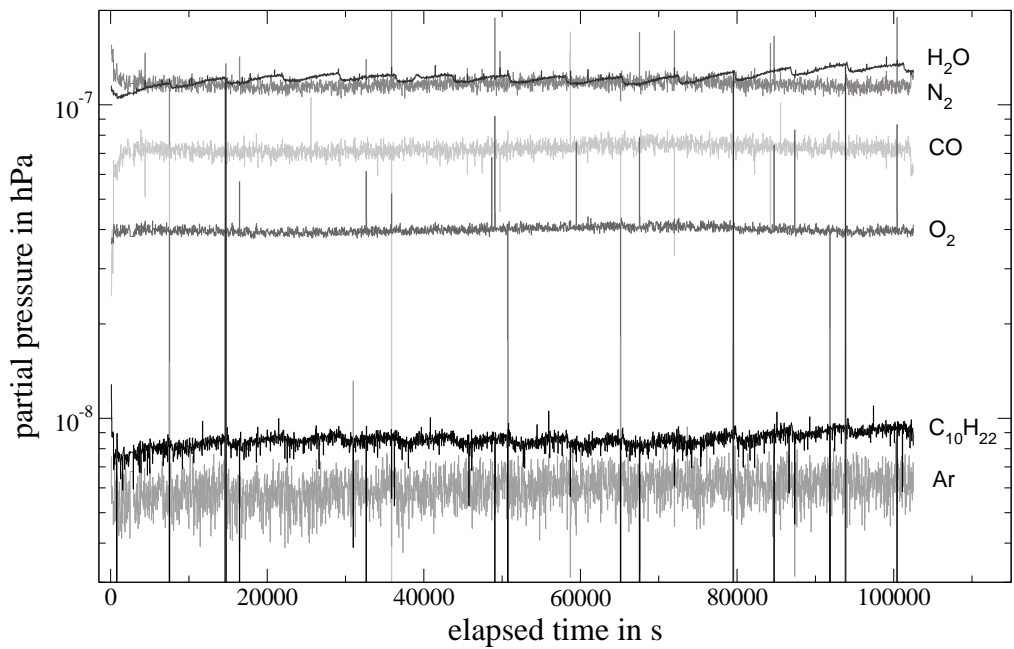}} \par}

\caption{\label{fig:PartD.Decan}Partial pressures during decane injection}
\end{figure}
 Fig.~\ref{fig:PartD.Decan} shows the recorded partial pressures during a measurement
with decane injection controlled by the regulator. 
\begin{figure}[!htbp]
{\par\centering \resizebox*{!}{0.9\textheight}{\includegraphics{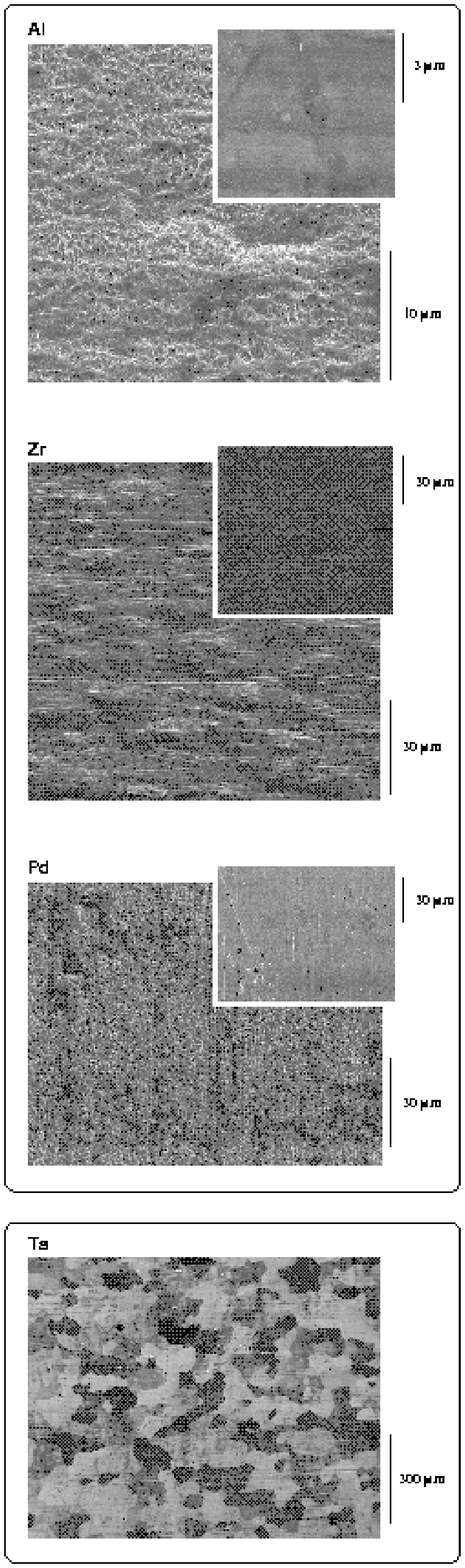}} \par}

\caption{\label{fig:emic}Scanning electron microscopic pictures of target surfaces
hit by the beam. The inserts (upper right) display areas not hit by the beam.}
\end{figure}
 The RGA was not calibrated since we are only interested in the relative fraction
of the residual gases. The partial pressures were calculated from the measured
partial currents as usual \cite{hanlon80} by solution of the linear equation
system with the cracking pattern matrix \cite{cornu66,massenspek} and singular
value decomposition. Even with the cryogenic trap water is the main constituent.
Its pressure changes with the refill periods of the trap in the target chamber.
Accordingly, the pressure of decane is adapted in order to retain the ratio.
The outliers in the curves originated from mismeasurements of the RGA which
are filtered out by the regulator program. Most interesting is the considerable
pressure of carbon monoxide that is almost twice as high as the one of oxygen
and on the other hand below the detection threshold without the beam. Although
\( ^{12}\mathrm{C}^{16}\mathrm{O} \) has no own molecular line in the mass
spectrum but needs to share it with \( ^{14}\mathrm{N}_{2} \) its fraction
is significant. This is a direct proof for the assumed chemical reaction at
the surface. The relative high fraction of argon is a holdover from the ventilation
of the chamber for what dry argon was used.

Once the gas mixture is offside the equilibrium one of the two possible layers
develops which proved to be irreversible. Further experiments evinced that the
equilibrium ratio is strongly beam energy dependent especially for the low energies.
Therefore the time consuming procedure of framing the equilibrium had to be
repeated. At \( 10\, \kilo \electronvolt  \) the C/O-ratio needs already to
be \( 1.6 \) in order to prevent oxidation. For \( 8\, \kilo \electronvolt  \)
even the maximal ratio of \( 3.0 \) for the outer regulating circle was not
sufficient. Only a pressure of \( 8\cdot 10^{-5}\, \hecto \pascal  \) could
prevent oxidation. This behaviour bases on the fact that the effect of the ion
beam on the target surface depends on the linear energy transfer into the target
which changes with the beam energy. The lower the projectile energy the higher
is the surface effect \cite{ensinger97}. This implies an ion flux dependency
incidentally which we could also observe. That is why we used fixed beam currents.
For zirconium the situation was even worse: At \( 25\, \kilo \electronvolt  \)
yet with the mounted gas conduit zirconium oxide formation could only be prevented
at high vapour pressures. On the other side for tantalum moderate pressures
were sufficient to counteract the oxidation. The equilibrium pressure was in
the interval from \( 3.0-4.0\cdot 10^{-6}\, \hecto \pascal  \) and showed a
comparatively very low energy dependence. This clear target material dependence
reflects the different interaction potentials and is in accord with the statements
of the previous section.

Albeit other processes interfere. More reactive metals i.e.~with lower electron
negativity oxidize more readily wherefore a higher decane pressure is required.
They also form chemically stronger bond metalhydrides whereas less reactive
metals merely allow for a segregation of hydrogen. Those reactive metals change
their crystal structure while forming the metalhydride. If the hydration precedes
not in a thermal equilibrium with low rates the material cannot compensate the
tension of the recrystallization process and bursts; embrittlement occurs. Since
deuteron implantation is far off the thermal equilibrium embrittlement is a
hardly avoidable concomitant phenomenon for reactive metals. Such greatly enlarges
the total surface while fractalizing it and enables the build up of thick metal
oxide layers. Fig.~\ref{fig:emic} contains scanning electron microscopic pictures
of some targets displaying areas inside the beam spot; the small inset pictures
are from areas outside the beam spot for comparison. Aluminium develops a sponge-like
coarse structure. It builds up a comparatively strong covalent bond to hydrogen
while transforming from the cubic face centered structure of the pure element
to the hexagonal structure of the hydride AlH\( _{3} \). Such definitely evokes
a lot of stress in the material. Zirconium traverses not just one but several
phase transitions with according changes of the crystal structure during the
increase of the hydrogen density \cite{beckM1}. The embrittlement is on a smaller
scale than for aluminium. The surface has a hill-like topography because the
crystallites in the polycrystalline material may have different hydrogen concentrations
and be in different phases with appendant crystal structures. The most reactive
metals are in the groups I and II of the periodic table. They developed the
strongest signs of embrittlement even on a macroscopic visible scale, e.g.~dust
particles crumbled from a strontium target, the thickness of a natrium target
grew considerably. For them the injection of decane would not help. On the other
side tantalum is almost a noble metal howbeit it can form a metallic bond to
hydrogen in contrast to noble metals which only segregate hydrogen with low
densities. It just stretches its lattice dimensions in order to accommodate
hydrogen in interstitial positions reffering to the original crystal structure.
Therefore the tension is low and can easily be compensated. Due to its noble
character the surface interaction potential of water favours the oxidation not
as much as for the more reactive metals why moderate decane pressures are sufficient.
The surface topography does not change. The spotted picture comes from beginning
layer formation and will be explained shortly. Palladium transmutes from cubic
face centered to cubic space centered while being hydrated. The surface shows
similarities to the case of aluminium. For Pd also a moderate decane pressure
was enough to prevent oxidation. Therefore it is an example that embrittlement
only promotes the metal oxide build up if the interaction potential already
favours dissociative chemisorption of water.

We carried out two experimental campaigns in oder to study the electron screening
effect. The results of the first campaign have been reported in previous publications
\cite{volos98,europhys01}. The second campaign had had the purpose to compact
the datapoints, heighten precission and to cover larger shares of the metals
in the periodic system. Therefore the experimental set-up including the accelerator
was completely renovated and revised in order to increase the beam current and
the reliability for long term measurements. The vacuum could also be improved
by almost an oder of magnitude. The principal set-up of the campaign~I is basically
the same as that of the campaign~II save of the gas injection and analyzing
equipment. With this new set-up only implantation curves of the type of the
lowest curve in Fig.~\ref{fig:AlImplant} were obtained which results from oxidation
where no screening can be observed. Water and with it insufficient vacuum was
soon identified as the reason. On the other hand the EDX-Analysis of the remaining
targets of the first campaign proofed that there was no additional metal oxide
layer. It raises the question how this could be possible with a worse vacuum.
Therefore the second campaign needed to change direction in order to investigate
this phenomenon and to reproduce the previous results. The carbon oxygen reaction
is the mechanism that keeps the surface clean. The higher gas load led to a
saturation of the cryogenic trap. Therefore backstreaming pumping oil is not
trapped as effective as in a better vacuum. The equilibrium where the target
surface keeps clean is much broader and therefore much less energy dependent
than for decane. It was hence adequate for the whole energy interval for most
metals. Though the measurement of the screening for strontium became difficult
and was hardly possible for lithium which are both very reactive metals the
later more than the first. The larger width of the equilibrium for pumping oil
is caused by the higher physisorption energy of the pumping oil because of the
greater molecular mass. Recalling that the van-der-Waals interaction of the
alcanes grows with every CH\( _{2,3} \)-group one can adopt that the physisorption
energy \( E_{d} \) of the pumping oil is \( 3.5 \)~times greater than that
of decane since decane has \( 10 \) carbon atoms and the pumping oil at least
\( 35 \). The hollow in the Lennard-Jones potential is accordingly deeper and
wider. The probability for desorption per second \( \nu  \) can be expressed
as \cite{henzler91} 
\begin{equation}
\label{eq:desorp}
\nu =\nu _{0}\exp \left( -\frac{E_{d}}{kT}\right) \: .
\end{equation}
 Thus, it becomes immediately clear that the desorption probability of decane
is \( 12 \)~times higher than that of the pumping oil for any temperature.
Furthermore, the change of the desorption probability \( \frac{d\nu }{dT} \)
is considerably higher for decane than for the pumping oil in the temperature
interval where desorption is still small enough in order to enable the surface
reaction. This contemplation concerns the dependency of the surface temperature
but is analogically applicable to straight beam induced desorption. In such
way the differences to the first campaign and its results are explained.

Since the equilibrium of decane is so narrow we abbreviated the total time of
irradiation of the measurements in order to obviate a drift off with subsequent
layer formation. This means also that the targets are not implanted up to the
stoichiometric ratios; so the depth distribution of the density may not be homogeneous
over the range of the ions.

\subsection{\label{sec:model}Modelling of layer effects}

In the case of an oxidation or carbon layer the presupposition that the depth
distribution of the deuterons in the target is homogeneous within the maximum
ion range is so no longer valid. This has consequences for the analysis of the
experimental results. Due to the alternating deuteron density at the target
surface the underlieing model for the data analysis needs to be modified which
means particularly that the single fitting parameter \( U_{e} \) is not enough
any more in order to describe \( F_{\mathrm{norm}} \) as obtained from the
experiment by (\ref{eq:ytab}.AB7). 
\begin{figure}[!htbp]
\[\resizebox*{\columnwidth}{!}{\input{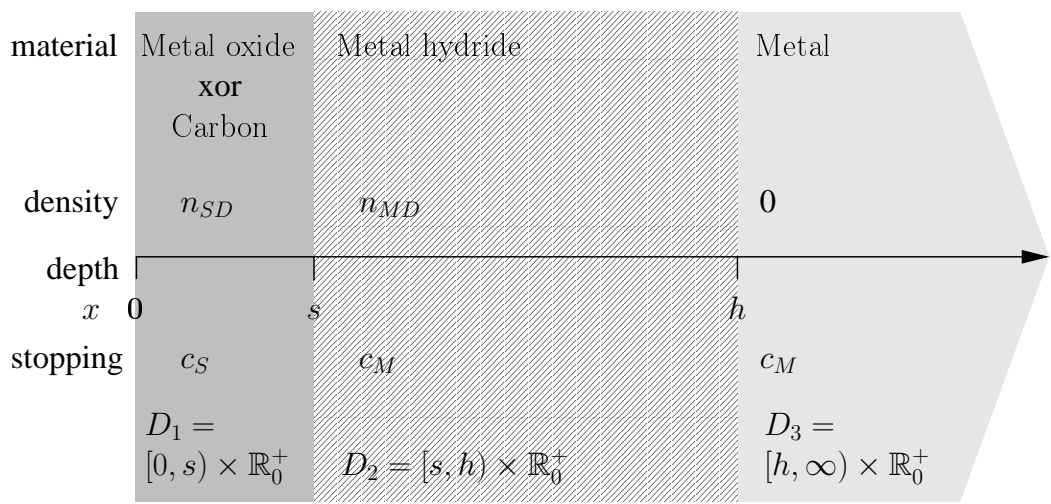}}\]

\caption{\label{fig:schichtstrukt}Model of the depth structure of the target.}
\end{figure}
 Fig.~\ref{fig:schichtstrukt} illustrates the modified model. The target is
made up of three layers. A surface layer consisting of either metal oxide or
carbon, a layer of metal hydride and the bulk of the metal. The model is based
on the following premises: The deuteron density in the metal oxide and in the
bulk of the metal is in comparison negligible and can be set to zero. The deuteron
density in carbon can achieve high values but there is no screening, so \( U_{e}=0 \).
There are sharp borders between these three layers. The deuteron density changes
correspondingly discontinuously. The later is not true for a real target, of
course where there is a smooth transition. However, the large energy steps of
the datapoints and a finite accuracy do not allow to resolve such a fine structure
of the experimental yield function. Starting with the definition of \( F_{\mathrm{norm}} \)
(\ref{eq:ytab}.AB7) coming back to the definition of the reduced yield (\ref{eq:ytab}.AB5)
a new expression for the theoretical yield function different from Eq.~(\ref{eq:ytab}.D6)
needs to be constructed. In Eq.~(\ref{eq:ytab}.BC1) the deuteron density is
now given by a piecewise defined function as in Fig.~\ref{fig:schichtstrukt}.
The energy substitution is performed with the stopping power differential equation
\begin{equation}
\label{eq:stoppingDGL}
\frac{dE}{dx}=\left\{ \begin{array}{cl}
-\left( c_{S}+\frac{n_{SD}}{n_{D}}c_{D}\right) E^{r} & ,\, D_{1}\\
-\left( c_{M}+\frac{n_{MD}}{n_{D}}c_{D}\right) E^{r} & ,\, D_{2}\\
-c_{M}E^{r} & ,\, D_{3}
\end{array}\right. 
\end{equation}
 It is now only piecewise continuous and defined on the domain \( D_{1}\cup D_{2}\cup D_{3} \)
with the belonging stopping power coefficients as in Fig.~\ref{fig:schichtstrukt}.
If \( r \) is set to \( \frac{1}{2} \) the generic continuous initial value
problem with the initial condition \( E\left( 0\right) =E_{0} \) has a handy
solution as a parabola in a vertex-like form 
\begin{equation}
\label{eq:stoppingFKT}
E\left( x\right) =\left( \frac{1}{2}c\cdot x-\sqrt{E_{0}}\right) ^{2}\: ,\quad x\in \left[ 0,\frac{2}{c}\sqrt{E_{0}}\right] 
\end{equation}
 describing the changes of the projectile energy within a given layer. Correspondingly
the yield integral can be splitted at the interval borders in three addends
where the last addend vanishes since the deuteron density in the metal, i.e.~the
third domain \( D_{3} \), is zero: 
\begin{eqnarray}
Y\left( E_{\mathrm{d}}\right)  & = & \frac{n_{SD}}{c_{S}+\frac{n_{SD}}{n_{D}}c_{D}}\int\limits _{E\left( s\right) }^{E_{\mathrm{d}}}\frac{\sigma \left( E\right) }{\sqrt{E}}dE\nonumber \\
 &  & +\frac{n_{MD}}{c_{M}+\frac{n_{MD}}{n_{D}}c_{D}}\int\limits _{E\left( h\right) }^{E\left( s\right) }\frac{\sigma \left( E+2U_{e}\right) }{\sqrt{E}}dE\label{eq:yieldlayer} 
\end{eqnarray}
 This can be inserted into the definitions of \( y \) and \( F_{\mathrm{norm}} \)

\begin{eqnarray}
F_{\mathrm{norm}}\left( E_{\mathrm{d}}\right)  & := & \frac{y\left( E_{\mathrm{d}}\right) }{y\left( E_{\mathrm{d},0}\right) }\nonumber \\
 & = & \frac{\overbrace{\frac{n_{SD}}{c_{S}+\frac{n_{SD}}{n_{D}}c_{D}}}^{\kappa _{S}}\frac{\int _{E\left( s\right) }^{E_{\mathrm{d}}}\frac{\sigma \left( E\right) }{\sqrt{E}}dE}{\int ^{E_{\mathrm{d}}}_{0}\frac{\sigma (E)}{\sqrt{E}}dE}+\overbrace{\frac{n_{MD}}{c_{M}+\frac{n_{MD}}{n_{D}}c_{D}}}^{\kappa _{M}}\frac{\int _{E\left( h\right) }^{E\left( s\right) }\frac{\sigma \left( E+2U_{e}\right) }{\sqrt{E}}dE}{\int ^{E_{\mathrm{d}}}_{0}\frac{\sigma (E)}{\sqrt{E}}dE}}{\underbrace{\frac{n_{SD}}{c_{S}+\frac{n_{SD}}{n_{D}}c_{D}}}_{\kappa _{S}}\frac{\int _{E_{0}\left( s\right) }^{E_{\mathrm{d},0}}\frac{\sigma \left( E\right) }{\sqrt{E}}dE}{\int ^{E_{\mathrm{d},0}}_{0}\frac{\sigma (E)}{\sqrt{E}}dE}+\underbrace{\frac{n_{MD}}{c_{M}+\frac{n_{MD}}{n_{D}}c_{D}}}_{\kappa _{M}}\frac{\int _{E_{0}\left( h\right) }^{E_{0}\left( s\right) }\frac{\sigma \left( E+2U_{e}\right) }{\sqrt{E}}dE}{\int ^{E_{\mathrm{d},0}}_{0}\frac{\sigma (E)}{\sqrt{E}}dE}}\label{eq:FnormLayer} \\
 & = & \frac{\overbrace{\frac{\kappa _{S}}{\kappa _{M}}}^{K}\frac{\int _{E\left( s\right) }^{E_{\mathrm{d}}}\frac{\sigma \left( E\right) }{\sqrt{E}}dE}{\int ^{E_{\mathrm{d}}}_{0}\frac{\sigma (E)}{\sqrt{E}}dE}+\frac{\int _{E\left( h\right) }^{E\left( s\right) }\frac{\sigma \left( E+2U_{e}\right) }{\sqrt{E}}dE}{\int ^{E_{\mathrm{d}}}_{0}\frac{\sigma (E)}{\sqrt{E}}dE}}{\underbrace{\frac{\kappa _{S}}{\kappa _{M}}}_{K}\frac{\int _{E_{0}\left( s\right) }^{E_{\mathrm{d},0}}\frac{\sigma \left( E\right) }{\sqrt{E}}dE}{\int ^{E_{\mathrm{d},0}}_{0}\frac{\sigma (E)}{\sqrt{E}}dE}+\frac{\int _{E_{0}\left( h\right) }^{E_{0}\left( s\right) }\frac{\sigma \left( E+2U_{e}\right) }{\sqrt{E}}dE}{\int ^{E_{\mathrm{d},0}}_{0}\frac{\sigma (E)}{\sqrt{E}}dE}}\, .\label{eq:Fnorm4P} 
\end{eqnarray}

In the last step (\ref{eq:FnormLayer}) is expanded by \( \frac{1}{\kappa _{M}}/\frac{1}{\kappa _{M}}=1 \).
The energies \( E\left( s\right)  \) and \( E\left( h\right)  \) at the borders
are calculated with the solution of the differential equation (\ref{eq:stoppingFKT})
in the current domain and the beam energy \( E_{d} \) to start from. 
\begin{eqnarray}
E\left( s\right)  & = & \bigg (\underbrace{\frac{1}{2}\left( c_{S}+\frac{n_{SD}}{n_{D}}c_{D}\right) \cdot s}_{\xi _{S}}-\sqrt{E_{\mathrm{d}}}\bigg )^{2}\label{m:schicht.Es} \\
E\left( h\right)  & = & \bigg (\underbrace{\frac{1}{2}\left( c_{M}+\frac{n_{MD}}{n_{D}}c_{D}\right) \cdot h}_{\xi _{M}}-\sqrt{E\left( s\right) }\bigg )^{2}\label{m:schicht.Eh} 
\end{eqnarray}
 \( E_{\mathrm{d},0} \) is the usual normalization energy of \( 25\, \kilo \electronvolt  \).
The function \( E_{0}\left( x\right)  \) distinguishes from \( E\left( x\right)  \)
only in the initial value \( E_{\mathrm{d},0} \) and Eq.~(\ref{m:schicht.Es}),
(\ref{m:schicht.Eh}) can be used correspondingly. The terms that contain the
dependencies from the densities and stopping power coefficients in Eq.~(\ref{eq:Fnorm4P}),
(\ref{m:schicht.Es}), (\ref{m:schicht.Eh}) are collected in the three parameters
\( K \), \( \xi _{S} \), and \( \xi _{M} \). Therefore once again, the analysis
is independent of the stopping power coefficients and the actual deuteron density.
\( K \) has the meaning of a weighting factor which describes the contribution
to the yield of the surface layer in comparison to the metal hydride and is
essentially determined by the ratio of the deuteron densities. For a metal oxide
layer it is set to zero since \( n_{SD}=0 \). The \( \xi  \)'s are energetic
lengths. With the stopping power coefficients and the densities the thickness
of the layers could be calculated from them. Practically only estimates can
be inferred since the densities in the layers are unknown and can change. Therefore
the model has now a quadrupel of four parameters: 
\begin{equation}
\label{eq:quadrupel}
\left\langle U_{e},K,\xi _{S},\xi _{M}\right\rangle 
\end{equation}

The model already provides some interesting insights. 
\begin{figure}[!htbp]
\resizebox*{!}{\textheight}{\input{fig10.pstex_t}}

\caption{\label{fig:schicht}Simulation of layer effects.}
\end{figure}
 Fig.~\ref{fig:schicht} shows some selected cuts out of the hyper area. The
plot (A) exhibits the effect of a metal oxide layer. The raise of \( F_{\mathrm{norm}} \)
at the lower energies quickly flattens with increasing thickness of the layer
and eventually reverses. This is exact the situation of the experimentally observed
reduced yield discontinuities in opposite direction for oxidized targets. Setting
\( c_{S} \) to the value for tantalum \( \xi _{S}=0.2\, \sqrt{\kilo \electronvolt } \)
where the former raise has already been completely vanished would correspondend
to a thickness of only \( 15\, \nano \meter  \). Thinner metal oxide layers
diminish the observed enhancement already considerably. A pure carbon layer
would have the same effect. But carbon is readily implanted by the beam with
deuterons to densities above \( 10^{22}\, \centi \meter \rpcubed  \). Therefore
the densities in carbon are comparable to the metal hydrides. Therefore in (B)
\( K \) is set to 1. The behaviour of \( F_{\mathrm{norm}} \) is similar to
(A) except that it does not drop below 1 since \( K\nless 1 \). On the other
hand if \( K>1 \) in a rather thick carbon layer, an enhancement could be artificially
generated because of the higher deuteron density in the carbon layer in comparison
to the metal. This is tested in (C), where additionally the screening energy
in the metal is set to zero. As expected, \( F_{\mathrm{norm}} \) raises indeed
with increasing \( K \), though the shape is different to the screening case.
The slope is less steep. This improves when the thickness of the carbon layer
becomes smaller. An inhomogeneous depth distribution of the deuterons causes
a considerable increase of \( F_{\mathrm{norm}} \) towards low energies while
the deuterated zone of the metal is becoming thinner (D). This additional rise
can only be distinguished from screening by the simultaneously amplified drop
at higher energies. That is why measurements at higher energies are still important.
Recapitulating, a carbon layer and/or an inhomogeneous depth distribution of
the deuteron density leads to an artificial increase of the enhancement which
might be erroneously attributed to the screening effect if no care is taken.

It should be noted that the thickness parameters of the layers in Eq.~(\ref{m:schicht.Es})
and Eq.~(\ref{m:schicht.Eh}) are not identical with the actual thickness of
the layers but are the way length that has been traversed by the ions. It is,
hence, longer because of multiple scattering. It is obvious that the difference
is of minor influence if the range of the ions is high in comparison to the
thickness. If, however, they are comparable the difference will be significant.
Such has been investigated quantitatively by a Monte-Carlo simulation of the
multiple scattering and stopping processes with a code adapted from Ref.~\cite{biller98}.
For the example of a carbon layer it turned out that for a thickness of \( 5\, \nano \meter  \)
there is no significant difference. Whereas for \( 40\, \nano \meter  \) the
difference is already a couple of \kilo\electronvolt. For \( E_{d}=5\, \kilo \electronvolt  \)
Eq.~(\ref{m:schicht.Es}) underestimates the stopping by \( 50\% \) albeit
such a thick layer levels all signs of the screening effect. Therefore this
is not relevant for surface layers. On the other hand the thickness of the metal
hydride zone is comparable to the range of the ions. But because the junction
between the metal hydride zone and the undeuterated metal is not discrete in
any case this is not relevant either.

The topics of Sec.~\ref{sec:anexpproc} and \ref{sec:surfreactlayer} are treated
of in more detail in Ref.~\cite{dis}.

\subsubsection{\label{sec:xlowenergy}Extreme low energy correction}

For very low beam energies \( U_{e}\approx E_{\mathrm{CM}} \) the simple picture
of the screening effect which attributes the screening energy to the kinetic
energy of the projectile as in Ref.~\cite{assenbaum87} does not hold any more.
The S-factor parameterization of the cross-section can be written as 
\begin{eqnarray}
\sigma \left( E_{\mathrm{CM}}\right)  & = & S\left( E_{\mathrm{CM}}\right) \underbrace{\frac{1}{\sqrt{E_{\mathrm{CM}}}}}_{\sim \lambda \; (\mathrm{a})}\label{eq:SfacXtrm} \\
 &  & \underbrace{\frac{1}{\sqrt{E_{\mathrm{CM}}+U_{e}}}\exp \left( -2\pi \eta \left( E_{\mathrm{CM}}+U_{e}\right) \right) }_{\sim P_{0}\; (\mathrm{b})}\nonumber 
\end{eqnarray}
where the term on the right side has been splitted in a factor including the
wavelength dependency (\ref{eq:SfacXtrm}.a) and the factor belonging to Coulomb
barrier penetration (\ref{eq:SfacXtrm}.b). Only to the latter one the screening
energy should be added \cite{EPS02a}. If this correction was to be applied
to the data analysis, Eq.~(\ref{eq:ytab}.D3) would need to be changed likewise.

\section{\label{sec:results}Results}

\begin{figure}[!htbp]
\resizebox*{1\columnwidth}{!}{\includegraphics{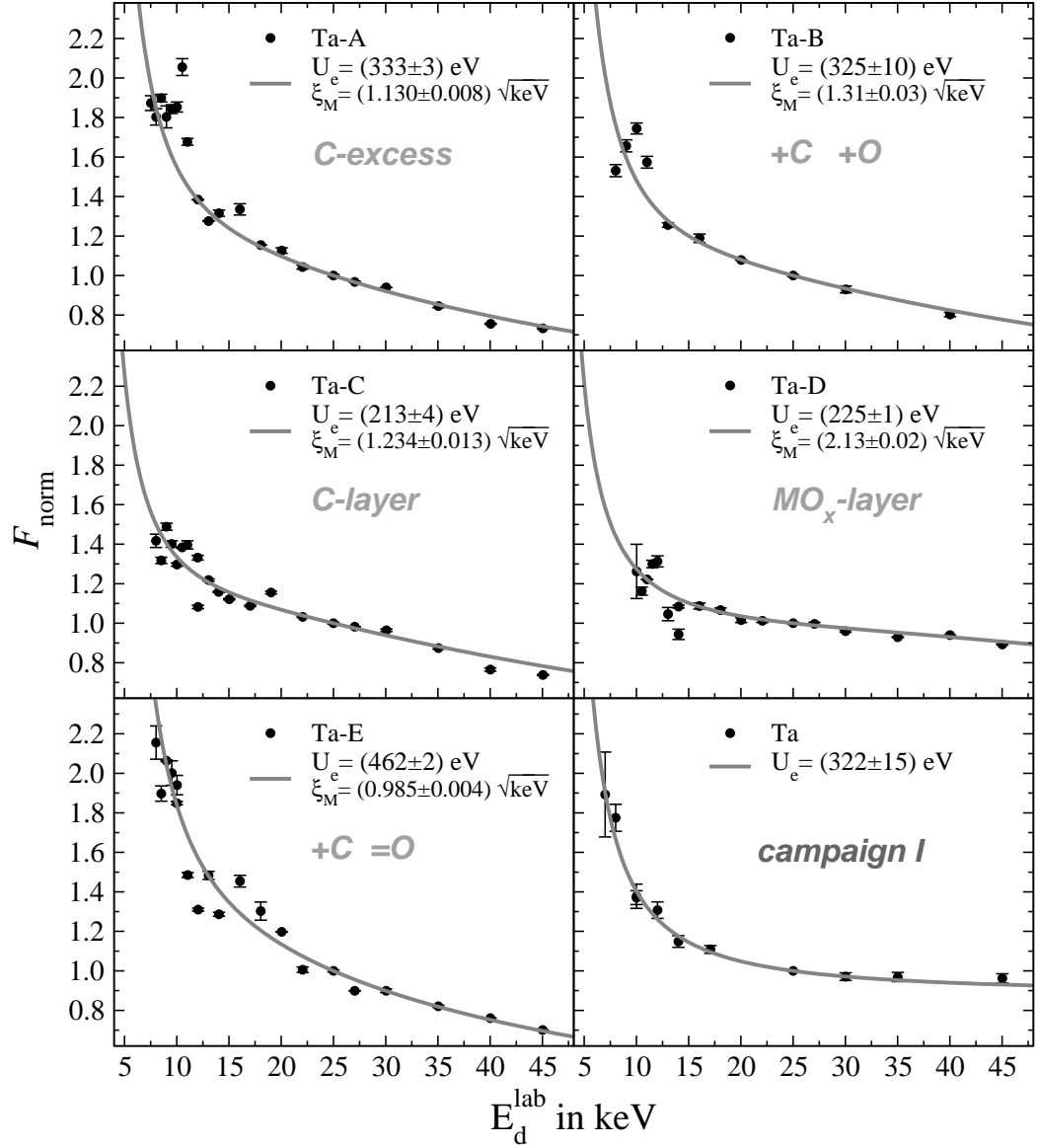}}

\caption{\label{fig:fituxim}Measurements on tantalum with different surface compositions.
Ta-A has a small C excess, Ta-B has slight C traces, Ta-C a thick C layer, Ta-D
a thick MO\protect\( _{x}\protect \) layer; Ta-E has slight C traces.}
\end{figure}

The ensured results for the lower limits of the campaign~I are listed in Table~\ref{tab:resultcampI}.
The corresponding plot for Tantalum resides in Fig.~\ref{fig:fituxim} (right,
bottom), for the other metals refer to Ref.~\cite{europhys01,dis}. In the second
column of the table the ratios of the deuterium number density to that of the
host metals are shown. Since the deuteron density can and does vary during a
measurement these values are estimated averages. The fourth column contains
the screening energy values obtained from the extreme low energy correction
from subsection \ref{sec:xlowenergy} with Eq.~(\ref{eq:SfacXtrm}) via Eq.~(\ref{eq:ytab}.D3).
In general this correction causes a slight reduction of the deduced screening
energy. The calculation has been performed with a different algorithm and illustrates
to some extent the influence of the algorithmic error which is a common part
of the error hierarchy in numerical mathematics. Anyway the correction affects
the result only at beam energies far below the current limit. 
\begin{table}[!htbp]

\caption{\label{tab:resultcampI}Screening energies from the campaign~I. \protect\( U_{e,x}\protect \)
is the screening energy applying the extreme low energy correction from (\ref{eq:SfacXtrm}).}
{\centering \begin{tabular}{llll}
\hline 
Metal&
MD\( _{x} \)&
\( U_{e} \) in \electronvolt&
\( U_{e,x} \) in \electronvolt\\
\hline 
Ta&
0.9&
322 \( \pm  \) 15&
300 \( \pm  \) 13\\
Zr&
2.1&
297 \( \pm  \) 8&
295 \( \pm  \) 6\\
Al&
0.8&
190 \( \pm  \) 15&
191 \( \pm  \) 12\\
\hline 
\end{tabular}\par}\end{table}

The campaign~II needed to focus on the investigation of the surface effects.
Fig.~\ref{fig:fituxim} shows the consequences on the thick target enhancement
factor \( F_{\mathrm{norm}} \) for Ta for a deliberate variation of the decane
infusion as an experimental counterpart to Fig.~\ref{fig:schicht} in the measurements
Ta-{[}A-E{]}. 
 The plot for Ta from the campaign~I is for comparison. Because of the reasons
stated in subsection \ref{sec:gasadmix} the total charge being implanted needed
to be limited. Hence, the deuterium density did not reach saturation. This has
two consequences that can be seen in the plots: First, a certain scattering
of the datapoints at the low energies which comes from small alterations in
the ion trajectories accompanying the adaption of the focus to the beam energy.
Then the beam spot shifts slightly and hits an area on the target with a different
deuteron density. Second, the depth distribution is not homogeneous any longer.
This leads to a drop of \( F_{\mathrm{norm}} \) at the higher energies and
a superelevation to the low energies like in Fig.~\ref{fig:schicht}.D. The
variation of the decane pressure was kept in the vicinity of the equilibrium
and caused changes in the composition of the target surfaces which were verified
with EDX-Analyses. Owing to the narrowness of the equilibrium, the targets showed
signs of beginning layer formation at least. The picture of an area inside the
beam spot of a Ta target in Fig.~\ref{fig:emic} exhibits distinct zones of
different brightness, i.e.\ emissivity for secondary electrons. The darker zones
have a greater carbon content than the brighter ones. With increasing overall
carbon load on the surface, the zones become darker and eventually vanish. This
phenomenon is known from ion beam assisted surface technology specifically for
thin film deposition \cite{ensinger97}. There are basically three modes for
the growth of thin films: Island growth, layer growth and the mixed case. Which
of them predominates depends on the affinity of the film material to the substrate.
The ion beam cares for the mobility of the adatoms at the surface e.g.\ by phonon
excitation. If the bond of the adatoms among themselves is stronger than to
the substrate atoms the adatoms move on the surface, join and form islands;
island growth will prevail. The islands will eventually grow together and form
a closed layer. The mutual bond of the carbon adatoms is stronger than that
to the atoms of the quasi noble metal tantalum entirely in antagonism to oxygen.
Therefore island growth is favoured. This also implies that the mobility of
the carbon atoms is high enough for it which also means that the surface is
sufficiently smooth and not destroyed by embrittlement.

\begin{figure}[!htbp]
{\par\centering \resizebox*{1\columnwidth}{!}{\includegraphics{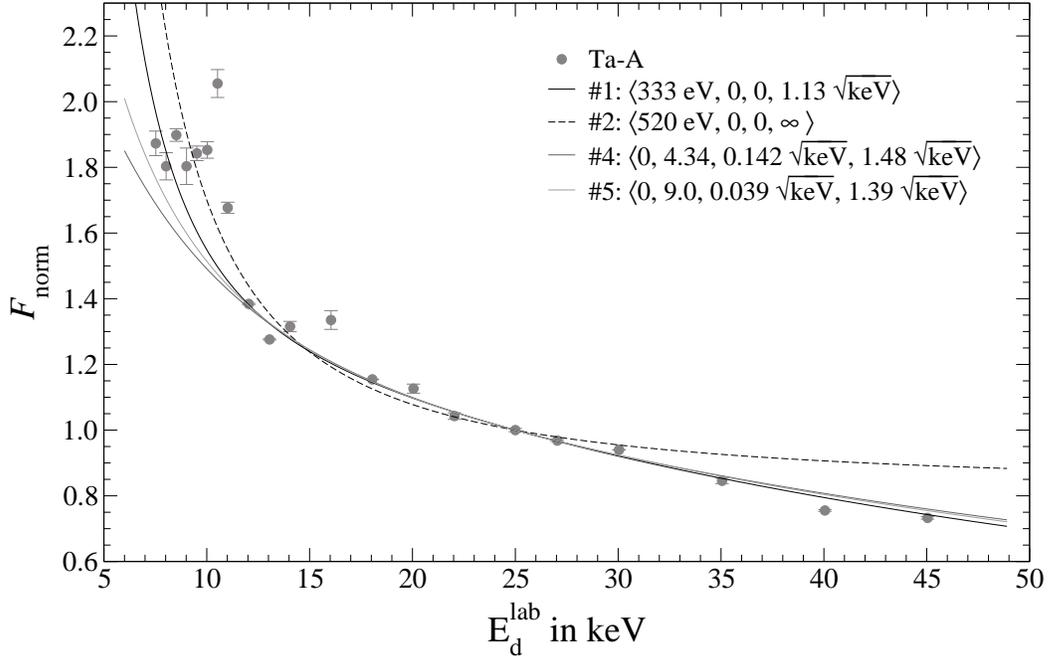}} \par}

\caption{\label{fig:FApp}Theoretical curves from different fit trials on the measurement
Ta-A, see table~\ref{tab:FApp}.}
\end{figure}

The curves in Fig.~\ref{fig:fituxim} are from fits with the 4-parameter model
of Sec.~\ref{sec:model} where the two parameters for the surface layer were
held fixed at zero. Since the \( \chi ^{2} \)-minimization problem is now nonlinear
it is not guaranteed that the iteration finds a global minimum and not only
local ones or even gets into an indefinite area, e.g.\ with a weak minimum.
In Table~\ref{tab:FApp} are some fit trials for different choices of parameters
with reasonable starting values. 
\begin{table*}[!htbp]

\caption{\label{tab:FApp}Results from fit trials. The gray printed numbers are prescribed
and fixed. \protect\( \widehat{\chi }^{2}:=\chi ^{2}/N\protect \), \emph{N}umber
of datapoints.}

\resizebox{\textwidth}{!}{
{\centering \begin{tabular}{rrddddddddd}
\hline 
&
No.&
\multicolumn{1}{l}{\( U_{e} \) {[}eV{]} }&
\multicolumn{1}{c}{\( \Delta U_{e} \)}&
\multicolumn{1}{c}{\( K \) }&
\multicolumn{1}{c}{\( \Delta K \)}&
\multicolumn{1}{c}{\( \xi _{S} \) {[}\( \sqrt{\textrm{keV}} \){]}}&
\multicolumn{1}{c}{\( \Delta \xi _{S} \)}&
\multicolumn{1}{c}{\( \xi _{M} \) {[}\( \sqrt{\textrm{keV}} \){]}}&
\multicolumn{1}{c}{\( \Delta \xi _{M} \)}&
\multicolumn{1}{c}{\( \widehat{\chi }^{2} \)}\\
\hline 
\textbf{Ta-A}&
1&
333&
3&
\gr{0} &
&
\gr{0}&
&
1.130&
0.008&
130.006\\
&
2&
520&
1&
\gr{0} &
&
\gr{0}&
&
\gr{\infty}&
&
539.136\\
&
3&
436&
413&
-0.26&
2.86&
0.05&
0.24&
1.04&
0.25&
129.636\\
&
4&
\gr{0} &
&
4.34&
0.14&
0.14&
0.02&
1.48&
0.04&
140.872\\
&
5&
\gr{0} &
&
9.0&
3.3&
0.04&
0.02&
1.39&
0.02&
135.660\\
\hline 
\textbf{Ta-C}&
6&
213&
4&
\gr{0} &
&
\gr{0}&
&
1.234&
0.013&
79.789\\
&
7&
326&
58&
\gr{1} &
&
0.12&
0.04&
1.20&
0.02&
84.149\\
\hline 
\textbf{Ta-D}&
8&
225&
1&
\gr{0} &
&
\gr{0}&
&
2.13&
0.02&
19.031\\
&
9&
234&
1&
\gr{0} &
&
\gr{0}&
&
\gr{\infty}&
&
53.219\\
&
10&
433&
130&
\gr{0} &
&
0.09&
0.06&
1.91&
0.13&
18.919\\
\hline 
\textbf{Ta-E}&
11&
462&
2&
\gr{0} &
&
\gr{0}&
&
0.985&
0.004&
130.785\\
&
12&
748&
1&
\gr{0} &
&
\gr{0}&
&
\gr{\infty}&
&
4850.515\\
\hline 
\end{tabular}\par}
}
\end{table*}
 The fits were performed with the Levenberg-Marquardt algorithm utilizing numerical
integration and differentiation procedures for the computation of the model
function and its gradient. The gray printed numbers are prescribed to the fitting
program and held fixed during runtime. Line~1 is the 2-parameter fit of Fig.~\ref{fig:fituxim}.
The plot of Ta-A is magnified in Fig.~\ref{fig:FApp} together with curves for
other fit attempts of Table~\ref{tab:FApp}. 
 The dashed curve in the figure is the 1-parameter fit \#2 merely for \( U_{e} \)
just like for the campaign~I. The curve shows a clearly increasing deviation
from the progression of the datapoints at the high energies above \( 30\, \kilo \electronvolt  \).
While attempting this the curve needed to be raised at the low energies, which
resulted in an inflated screening energy of \( 520\, \electronvolt  \). Such
behaviour is in accord with Fig.~\ref{fig:schicht}.D. A fit with all 4 parameters
is \#3 (see line 3 in Table~\ref{tab:FApp}). The belonging covariance matrix
has large off diagonal elements hence the parameters are strongly correlated.
The large errors of the fit parameters are another indication. The surface layer
parameters \( K \) and \( \xi _{S} \) equals to zero within their errors.
This does not mean that there is no surface layer but only that its effect is
below the resolution of the datapoints. In analogy to Fig.~\ref{fig:schicht}.C
it should be tested whether it is possible to describe the datapoints by the
layer parameters alone, i.e.\ the screening energy is fixed to zero. This is
done in \#4 where the result is from a first execution of the fitting procedure
which got into an indefinite area of weak convergence. The values after repetitive
iterations without reaching a minimum are in \#5. However, a tendency becomes
clear: \( K \) becomes still greater on the expense of \( \xi _{S} \) which
is reduced at the same time. Those values for the parameters -- particularly
\( K \) -- are physically not realistic. The associated curves can not follow
the slope at the low energies. Ta-C has a thick carbon layer. Therefore the
enhancement factors are considerably reduced like in Fig.~\ref{fig:schicht}.B
as well as the screening energy from the 2-parameter fit \#6 (line 6 in Table~\ref{tab:FApp}).
A 3-parameter fit with \( K \) set to 1 complying to a carbon layer yields
the values in \#7 now with a screening energy in agreement to the previous results.
Admittedly, it belongs to a weak minimum. By far the most total charge was implanted
at Ta-D under conditions where the formation of a metal oxide layer is favoured.
Accordingly, the thickness of the deuterated zone \( \xi _{M} \) is considerably
larger and the screening is inhibited by the layer (\#8). Suitably the 1-parameter
fit \#9 has a similar result since the high energy datapoints have no significant
drop, too. This time the 3-parameter fit \#10 found a local minimum with a surface
thickness of \( \xi _{S}=0.09\, \sqrt{\kilo \electronvolt } \) which is sufficient
to halve the enhancement factors (Fig.~\ref{fig:schicht}.A) and corresponds
to a thickness of about \( 7\, \nano \meter  \) on tantalum, although the three
parameters have a clear correlation as the errors indicate. The found screening
energy agrees with the larger value of Ta-E \#11. It has a lower carbon load
than Ta-A and the measured enhancement factors are higher. The 1-parameter fit
\#12 yields an even more unrealistic value than \#2. Because of the necessarily
insufficient implantation the 2-parameter fit is best suited since the scattering
of the datapoints seldom allows for a fit with more free parameters and then
only limited validity.

\section{\label{sec:discuss}Discussion}

Several possible (mostly minor) error sources have been already discussed in
Ref.~\cite{europhys01,nimb02} and won't be repeated here. Instead, the discussion
concentrates on the dominating error source that is the inhomogeneous deuteron
density distribution in the targets. We concluded that our previous values for
the screening energies represent lower limits of the real ones. As pointed out
in section \ref{sec:experiment} the actual energy of the projectiles is somewhat
lower than measured at the voltage divider because of a voltage drop inside
HF ion sources. Therefore the observed screening energy is also lowered by this
property.

A possible question is whether the low energy expression for the stopping power
function \( E^{\frac{1}{2}}\propto v \) is valid for the metal hydrides or
if there is a significant energy dependence of the superscript or for that matter
of the stopping power coefficients. In this case part of the enhancement could
be caused by this conceivable effect. Albeit the Lindhard theory has been experimentally
validated down to \( 1\, \kilo \electronvolt  \) \cite{paul91} measurements
at He and Ne gas showed a deviation at low energies \cite{golser91,schiefermueller93}.
This is due to the high excitation energies of both noble gases \cite{grande93,semrad86}.
In solids the velocity proportionality is fully confirmed for metals as well
as for insulators with wide band gaps like LiF, Al\( _{2} \)O\( _{3} \) and
SiO\( _{2} \) \cite{eder97,moller02,moller04}.

\begin{figure*}[!htbp]
{\par\centering \resizebox*{1\textwidth}{!}{\includegraphics{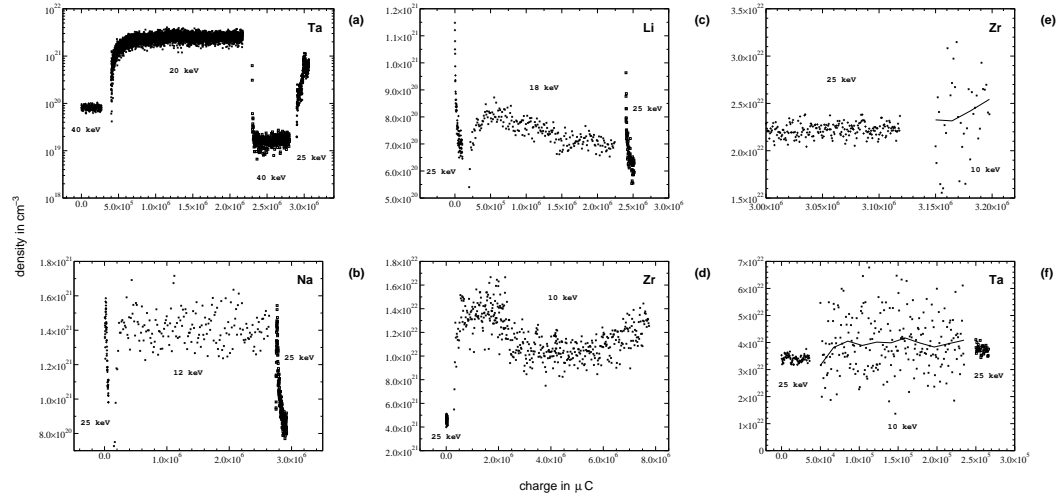}} \par}

\caption{\label{fig:counterex}Counter examples to the case of the screening enhancement
discontinuities in fig.~\ref{fig:expprocedure}. First case: Low hydrogen binding
ability of the host material (a); here because of heating of thin target foils.
Second case: Density profile changes in thick metal oxide layers (b-d). Third
case: Enhancement discontinuities are leveled off by medium thick layers at
high densities (e,f).}
\end{figure*}

According to the basic theoretical definition of the yield \( Y=\int ^{R}_{0}\left[ n\sigma \right] dx \)
(\ref{eq:ytab}.BC1), deviations in the observed yield and correspondingly in
the reduced yield have the two principal causes: changes in the deuteron density(-profile)
and modification of the cross-section probably by the screening effect merged
in the integrant product \( \left[ n\sigma \right]  \). As already pointed
out in section \ref{sec:advanalysis} and showed in Fig.~\ref{fig:expprocedure},
our analysis method allows us to discern between these two causes. Only the
direct reduced yield discontinuity at the change of the beam energy is interpreted
as a modification of the cross-section. The differential method enables the
recognition and rejection of measurements by observation of the progression
of \( y\left( q\right)  \) which could easily be misinterpreted as screening
if merely the total yields over an entire measurement are compared. Some representative
examples for such cases are collected in Fig.~\ref{fig:counterex} and elucidate
it. 
 The quantity on the vertical axis is herein the density as computed with Eq.~(\ref{m:red.nDichtey})
since it is more illustrative.

Subfig.~\ref{fig:counterex}.a demonstrates consequences of a perceptible heating
of the target. Since the beam energy is deposited inside the target in a volume
with a thickness \( <1\, \micro \meter  \) the thermal energy needs to be led
away otherwise this zone will be considerably heated. The far most effective
heat transport mechanism in solids is heat conduction by the free electrons
inside metals. Hence, we used thick targets (\( \cong 1\, \milli \meter  \))
where the thermal energy can be dissipated into the bulk of the material and
so the temperature is kept low. The target holders are equipped with channels
for coolant flow. However, the comparatively very high thermal resistance of
the joint between the holder and the mounted target allows only for a small
effect, at best. Therefore for our usual beam currents of some \( 10\, \micro \ampere  \)
the density remained essentially constant at the change of the beam energy.
Only long term changes were observed as described in Sec.~\ref{sec:gasadmix}.
The effects of heating were investigated using a thin Ta-foil (\( 7\, \micro \meter  \))
and higher currents where the energy dissipation is entirely insufficient and
hence the target heated. Then the 'saturation' densities are heavily dependent
on the beam energy and current and much lower than the stoichiometric ratio.
Albeit, for the same conditions, i.e.~beam energy, current, target thickness
and volume, the density returns to the same 'saturation' level. Now Subfig.~\ref{fig:counterex}.a
displays the development of the density during the progression of the different
measurements. When switching from \( 40\, \kilo \electronvolt  \) to \( 20\, \kilo \electronvolt  \)
the density starts with a reverse discontinuity and rapidly increases within
\( 10^{5}\, \micro \coulomb  \) and remains then constant. Switching back to
\( 40\, \kilo \electronvolt  \) the density decreases more rapidly and levels
off. Going to \( 25\, \kilo \electronvolt  \) once again the initial rapid
increase can be observed reaching a first plateau and then a second after reduction
of the current. Most metals form a chemical bond with the hydrogen whose strength
depends on the relative affinity. Therein the hydrogen will accumulate in the
volume of the beam retardation until all bonds are saturated in vicinity of
the stoichiometric ratio and then migrate into adjacent crystallites filling
them up etc. If, however, the temperature is too high in order to enable the
formation of the chemical bond the stopped hydrogen will diffuse through the
whole material promoted by the high temperature and may even be able to return
to the gas phase. The then attuning 'saturation' density is a labile equilibrium
state highly dependent on the surrounding conditions. The area of our target
holders and materials was large enough to accommodate 4 beam spots. Changes
between the beam positions were done by a vertical shift of the target holder
(Fig.~\ref{fig:aufbau}). Going to another spot position the initial implantation
curve starts at zero like in Fig.~\ref{fig:AlImplant} but not so at the Ta-foil.
There the density already starts at a value close to the previous spot sustaining
the explanation. On the foil were folds around the beam spots indicating the
thermal stress that the foil suffered. The fundamental reason for the described
density dynamics at low absolute density values is the enlarged mobility of
the deuterons inside the metal. It is here in the case of tantalum caused by
heating but will occur whenever the thermal energy of the deuterons is higher
than their chemical binding energy to the metal so that they can float free.
For metals with low ability to bind hydrogen, e.g.~the transition metals in
the groups 6A-8A and 1B, this is already fulfilled at room temperature. A test
with a thin Au foil showed a behaviour like for the Ta foil without the screening
enhancement discontinuities at a very low density of \( 3\cdot 10^{21}\, \centi \meter \rpcubed  \)
(\( x=0.05 \)). Besides indirect heating, the mobility of the deuterons in
target materials with low binding ability for deuterons is also promoted by
direct projectile scattering and phonon generation depending on the beam energy.

Due to the inability of metal oxides to form a chemical bond to the hydrogen
(Sec.~\ref{sec:surfeff}) the same as above applies to them and is additionally
aggravated by the fact that the metal oxides are not only electrical but also
thermal insulators so (beam)-heating is more effectual. Subfig.~\ref{fig:counterex}.b
illustrates the development of the densities in a Na target with a thick oxide
layer. Before the monitor measurement at \( 25\, \kilo \electronvolt  \) a
measurement at a low energy had been taken. The density quickly decreased then
at \( 25\, \kilo \electronvolt  \). Thereafter a measurement at \( 12\, \kilo \electronvolt  \)
was started. The density very quickly increased reaching a higher level than
at \( 25\, \kilo \electronvolt  \). But the discontinuity at the beginning
was in the wrong direction. The density for the sequencing monitor measurement
started once again at a high density which quickly decreased. The discontinuity
at the beginning was once again in the wrong direction. Therefore there is definitely
no screening. The discontinuities result in an enhancement factor \( F_{\mathrm{norm}}<1 \)
conforming with Fig.~\ref{fig:schicht}.A. The quick shifts in the densities
after the change of the beam energy going to a 'saturation' level originate
from a shift of the deuteron distribution depth profile in the metal oxide linked
to the different ranges of the ions. 
\begin{figure}[!htbp]
{\par\centering \resizebox*{0.9\columnwidth}{!}{\includegraphics{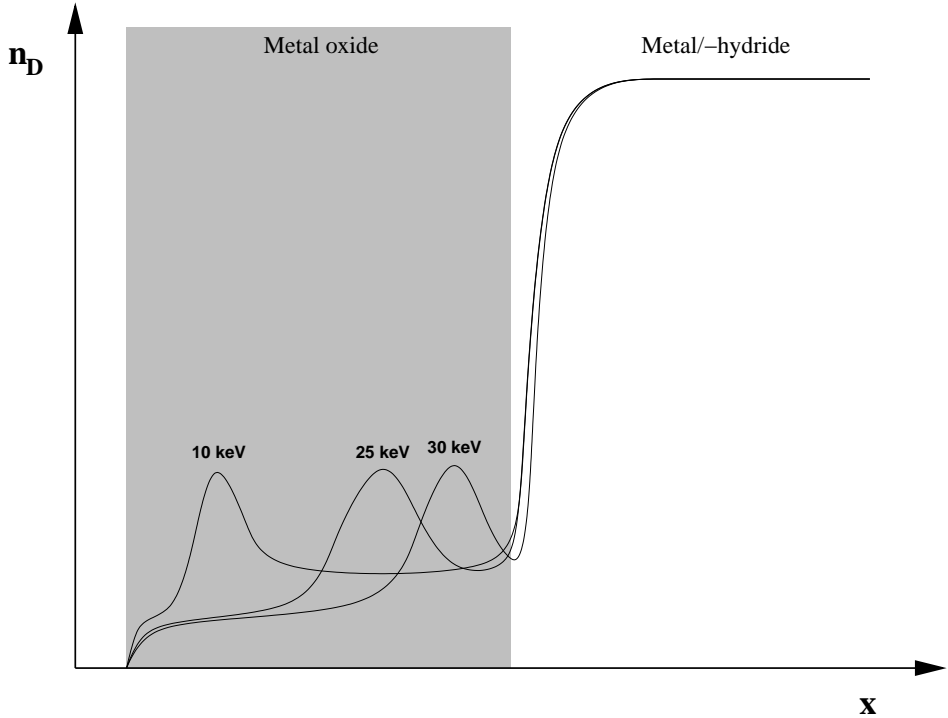}} \par}

\caption{\label{fig:DichteMoxid}Deuteron number density profiles in oxidized metals
at different energies.}
\end{figure}
 This is schematically sketched in Fig.~\ref{fig:DichteMoxid}. Since hydrogen
is only segregated in metal oxides the density is at least one order of magnitude
less than in the metal hydride below. Inside the metal oxide the density is
maximal around the range of the ions and the deuterons migrate from there. Between
the surface and the range the density is very low since the linear energy transfer
of the beam ions exceeds the low segregation energy by far. Therefore only a
labile diffusion equilibrium can attune. Beyond the range the density is higher
than before because the binding in segregation is sooner effective. Beginning
at a longer irradiation at \( 25\, \kilo \electronvolt  \) and then changing
to \( 30\, \kilo \electronvolt  \) the maximum of the density profile of \( 25\, \kilo \electronvolt  \)
is within the ion trajectory at \( 30\, \kilo \electronvolt  \). At first the
homogeneous average density calculated with Eq.~(\ref{m:red.nDichtey}) appears
accordingly distinctly higher since the ions possess a higher energy for augmented
reactions at this position. Concurrently the density maximum is dismantled by
the energy transfer and metamorphosed into the profile of \( 30\, \kilo \electronvolt  \)
while the average density is decreasing. Is then a measurement at \( 10\, \kilo \electronvolt  \)
performed the density within the deuteron range is very low but is swiftly augmented
by the stopped ions. Therefore the calculated average density is increasing.
After such a measurement had been carried out long enough the density beyond
the range is enlarged by segregation and a sequencing measurement at a higher
energy yields again a strongly enhanced average density which is quickly decreasing.
The same interplay can be observed in the case of Li in Subfig.~\ref{fig:counterex}.c
with a lower energy difference.

With decreasing energy difference between consecutive measurements the differences
in the density profiles become smaller but remain still significant. Subfig.~\ref{fig:counterex}.d
for Zr with a thick oxide layer is in direct contradistinction to fig.~\ref{fig:expprocedure}.
The calculated average densities are less than \( \frac{1}{10} \) in comparison
to Fig.~\ref{fig:expprocedure} and the characteristical quick increase at the
beginning of the \( 10\, \kilo \electronvolt  \) measurement is visible with
no screening discontinuity. Here should be emphasized the importance of the
beam adjustment procedure with the deflector offside the target as described
in Sec.~\ref{sec:experiment} for the observation of the quick changes at the
beginning of the measurements. In Subfig.~\ref{fig:counterex}.e the oxide layer
is thinner and the density not so much abated. Hence, there is no such quick
increase at \( 10\, \kilo \electronvolt  \). The line is calculated with a
larger step size thus smoothing the progression. The deuteron density for Ta
in Subfig.~\ref{fig:counterex}.f is at the stoichiometric ratio but there is
no discontinuity and no quick change when switching the energy. It complies
to a layer with a thickness of \( \xi _{S}=0.2\, \sqrt{\kilo \electronvolt } \)
in Fig.~\ref{fig:schicht}.A where the normalized enhancement factor equals
zero and which corresponds to a low thickness of \( 15\, \nano \meter  \) Ta.
Measurements which yielded results like they are represented in Fig.~\ref{fig:counterex}
were altogether discarded since the changes in the yield are only from the density
dynamics and not from screening. Summarizing, it is preferable to use thick
target disks at moderate temperatures with high densities. On the other side,
cooling a target to very deep temperatures would transform it into a cryogenic
trap accumulating water in thick layers on its surface prior to irradiation
promoting the oxidation. Paying attention to all the above discussed experimental
problems we can state that our results represent lower limits to the real screening
energy values. For further details refer to Ref.~\cite{dis}.

\section{Conclusion}

We alluded to experimental anomalies and pitfalls of the d+d reactions in metals
which distinguish them from common nuclear reaction measurements and make them
particularly difficult and error-prone. With respect to these special conjectures
we developed a differential data analysis method which gains the maximum information
from the raw data. It enables the on-line monitoring of the deuteron densities
and the observation of short time deuteron density profile changes. Only the
discontinuities of the reduced reaction yields at the change of the beam energies
can be attributed to the screening effect and not relative changes in the total
yield of the elapsed measurements. The method is independent of the unprecise
stopping power coefficients and the actual absolute value of the deuteron number
density in the targets. Since the measurements are taken at deuteron densities
in the vicinity of the stoichiometric ratio the yields cannot be sophisticated
by density dynamics. The measurements are further impaired by layer formation
under the beam irradiation supplied by the residual gas in the high vacuum system
used in nuclear physics. Dedicated experiments demonstrated how the composition
of the residual gas influences the layer formation and the measurements whereby
the chemical active fractions are water and carbon hydrides. These processes
depend on many difficult to control parameters. The experiments on tantalum
showed the effects on the results for the inferred screening energy utilizing
the model for the layer effects. In consequence our results for the screening
energies represent lower limits. The results from the tantalum experiments range
from \( 210-460\, \electronvolt  \) giving an imagination of the error in comparison
to the lower limit of \( 322\pm 15\, \electronvolt  \). The layer model shows
that a few ten atomic layers of different composition are enough in order to
obliterate the screening effect. Thicker metal oxide layers lead to low and
instable densities. The applied techniques allow for the recognition and rejection
of measurements with unwanted shifts in the density depth distribution profile
thus preventing the erroneous extraction of an artificial screening enhancement
in contrast to the standard analysis based on the total yield measurement used
by Ref.~\cite{yuki98,kasagi02,rolfs02b,rolfs03,rolfs04,rolfs05b,rolfs06}\footnote{%
Such will be discussed and substantiated in a forthcoming publication.
}. Those undesirable density profile changes occur in targets with low hydrogen
binding ability, like many of the transition metals, at elevated temperatures
and heterogeneous targets with metal oxide or carbon layers or different (relatively)
thin metal layers. Therefore aside from the sustained fact that there is a great
screening enhancement so far no further assertion can be made about possible
causes based on other known material properties because of the interference
with the chemical surface reactions. For a precise determination of the screening
energies ultra high vacuum systems with pressures well below \( 10^{-10}\hecto \pascal  \),
where only hydrogen and noble gases are in the residual gas, and equipped with
in-situ target diagnosis techniques are mandatory.


\begin{thebibliography}{10}
\expandafter\ifx\csname url\endcsname\relax
  \def\url#1{\texttt{#1}}\fi
\expandafter\ifx\csname urlprefix\endcsname\relax\def\urlprefix{URL }\fi

\bibitem{assenbaum87}
H.~J. Assenbaum, K.~Langanke, C.~Rolfs, Z. Phys. A~(327) (1987) 461--468.

\bibitem{salpeter54}
E.~E. Salpeter, Aust. J. Phys. 7 (1954) 373.

\bibitem{ichimaru93}
S.~Ichimaru, Rev. Mod. Phys. 65 (1993) 252.

\bibitem{volos98}
K.~Czerski, A.~Huke, P.~Heide, M.~Hoeft, G.~Ruprecht, in: N.~Prantzos,
  S.~Harissopulos (Eds.), Nuclei in the Cosmos V, Proceedings of the
  International Symposium on Nuclear Astrophysics, Editions Fronti{\`{e}}res,
  Volos, Greece, 1998, p. 152.

\bibitem{europhys01}
K.~Czerski, A.~Huke, A.~Biller, P.~Heide, M.~Hoeft, G.~Ruprecht, Europhys.
  Lett. 54~(4) (2001) 449--455.

\bibitem{dis}
A.~Huke, {D}ie {D}euteronen-{F}usionsreaktionen in {M}etallen, Ph.D. thesis,
  Technische Universität Berlin (2002).
\newline\urlprefix\url{http://edocs.tu-berlin.de/diss/2002/huke_armin.htm}

\bibitem{greife95}
U.~Greife, F.~Gorris, M.~Junker, C.~Rolfs, D.~Zahnow, Z. Phys. A~(351) (1995)
  107--112.

\bibitem{yuki98}
H.~Yuki, J.~Kasagi, A.~G. Lipson, T.~Ohtsuki, T.~Baba, T.~Noda, B.~F. Lyakhov,
  N.~Asami, JETP Lett. 68~(11) (1998) 823.

\bibitem{kasagi02}
J.~Kasagi, H.~Yuki, T.~Baba, T.~Noda, T.~Ohtsuki, A.~G. Lipson, J. Phys. Soc.
  Jpn. 71 (2002) 2281.

\bibitem{rolfs02}
F.~Raiola, et~al., Eur. Phys. J. A 13 (2002) 377.

\bibitem{rolfs02b}
F.~Raiola, et~al., Phys. Lett. B 547 (2002) 193.

\bibitem{rolfs03}
C.~Bonomo, et~al., Nucl. Phys. A 719 (2003) 37c.

\bibitem{rolfs04}
F.~Raiola, et~al., Eur. Phys. J. A 19 (2004) 283.

\bibitem{europhys04}
K.~Czerski, A.~Huke, P.~Heide, G.~Ruprecht, Europhys. Lett. 68~(3) (2004)
  363--369.

\bibitem{rolfs88}
C.~E. Rolfs, W.~S. Rodney, Cauldrons in the Cosmos, Theoretical Astrophysics,
  The University of Chicago Press, Chicago and London, 1988.

\bibitem{NPAII06b}
K.~Czerski, A.~Huke, P.~Heide, G.~Ruprecht, Eur. Phys. J. A 27~(S1) (2006)
  83--88.

\bibitem{kamke56}
D.~Kamke, Handbuch der Physik, Vol. XXXIII, Springer Verlag, Berlin, 1956, Ch.
  1. Elektronen- und Ionenquelle.

\bibitem{brown90}
R.~E. Brown, N.~Jarmie, Phys. Rev. C 41~(4) (1990) 1391.

\bibitem{mhydrides68}
W.~M. Mueller, J.~P. Blackledge, G.~G. Libowitz (Eds.), Metal Hydrides,
  Academic Press, New York, London, 1968.

\bibitem{lindhard54}
J.~Lindhard, Mat. Fys. Medd. Dan. Vid. Selsk. 28~(8).

\bibitem{lindhard61}
J.~Lindhard, M.~Scharff, Phys. Rev.~(124) (1961) 128.

\bibitem{lindhard63}
J.~Lindhard, M.~Scharff, H.~E. Schi{\o}tt, Mat. Fys. Medd. Dan. Vid. Selsk.
  33~(14).

\bibitem{ziegler77}
H.~Anderson, J.~F. Ziegler, The Stopping and Ranges of Ions in Matter, Vol.~3,
  Pergamon Press, New York, 1977.

\bibitem{bragg05}
R.~Bragg, Phil. Mag.~(10) (1905) 318.

\bibitem{abramowitz}
M.~Abramowitz, I.~A. Stegun, Handbook of Mathematical Functions, 9th Edition,
  Dover Publications, Inc., New York, 1970.

\bibitem{henzler91}
M.~Henzler, W.~Göpel, Oberflächenphysik des Festkörpers, B. G. Teubner,
  Stuttgart, 1991.

\bibitem{zangwill88}
A.~Zangwill, Physics at surfaces, Cambridge University Press, Cambridge, 1988.

\bibitem{chabal84}
Y.~J. Chabal, S.~B. Christmann, Phys. Rev. B~(29) (1984) 6974.

\bibitem{scheffler84}
M.~Scheffler, A.~M. Bradshaw, The electronic structure of adsorbed layers, in:
  D.~A. King, D.~P. Woodruff (Eds.), Adsorption on Solid Surfaces, Vol.~2 of
  The Chemical Physics of Solid Surfaces and Heterogeneous Catalysis, Elsevier,
  Amsterdam, 1984.

\bibitem{hering99}
W.~T. Hering, Angewandte Kernphysik, B. G. Teubner, Stuttgart, Leipzig, 1999.

\bibitem{hanlon80}
J.~F. O'Hanlon, A User's Guide to Vacuum Technology, John Wiley {\&} Sons, New
  York, 1980.

\bibitem{cornu66}
A.~Cornu, R.~Massot, Compilation of Mass Spectral Data, Heyden and Son Limited,
  London, 1966.

\bibitem{massenspek}
Leybold-Heraeus GmbH, Köln, Quadruvac Q 100 (1983).

\bibitem{ensinger97}
W.~Ensinger, Nucl. Instr. Meth. B~(127/128) (1997) 796.

\bibitem{beckM1}
R.~L. Beck, W.~M. Mueller, Zirconium hydrides and hafnium hydrides, in: Mueller
  et~al.  \cite{mhydrides68}, Ch.~7.

\bibitem{biller98}
A.~Biller, {E}influß der {V}ielfachstreuung auf {D}icktarget-{Y}ields in
  niederenergeti\-schen {K}ernreaktionen, Diplomarbeit, Institut für Atomare
  und Analytische Physik der Technischen Universität Berlin (1998).

\bibitem{EPS02a}
K.~Czerski, A.~Huke, P.~Heide, Nucl. Phys. A 719 (2002) 52c.

\bibitem{nimb02}
K.~Czerski, A.~Huke, P.~Heide, G.~Schiwietz, Nucl. Instr. Meth. B 193 (2002)
  183.

\bibitem{paul91}
H.~Paul, D.~Semrad, A.~Seilinger, Nucl. Instrum. Meth. B 61 (1991) 261.

\bibitem{golser91}
R.~Golser, D.~Semrad, Phys. Rev. Lett. 66 (1991) 1831.

\bibitem{schiefermueller93}
A.~Schiefermüller, R.~Golser, R.~Stohl, D.~Semrad, Phys. Rev. A 48 (1993) 4467.

\bibitem{grande93}
P.~L. Grande, G.~Schiwietz, Phys. Rev. A 47 (1993) 1119.

\bibitem{semrad86}
D.~Semrad, Phys. Rev. A 33 (1986) 1646.

\bibitem{eder97}
K.~Eder, D.~Semrad, P.~Bauer, R.~Golser, P.~Maier-Komor, F.~Aumayr,
  M.~Pe{\~{n}}alba, A.~Arnau, J.~M. Ugalde, P.~M. Echenique, Phys. Rev. Lett.
  79 (1997) 4112.

\bibitem{moller02}
S.~P. M{\o}ller, A.~Csete, T.~Ichioka, H.~Knudsen, U.~I. Uggerh{\o}j, H.~H.
  Andersen, Phys. Rev. Lett. 88 (2002) 193201.

\bibitem{moller04}
S.~P. M{\o}ller, A.~Csete, T.~Ichioka, H.~Knudsen, U.~I. Uggerh{\o}j, H.~H.
  Andersen, Phys. Rev. Lett. 93 (2004) 042502.

\bibitem{rolfs05b}
F.~Raiola, et~al., J. Phys. G 31 (2005) 1141.

\bibitem{rolfs06}
F.~Raiola, et~al., Eur. Phys. J. A 27~(S1) (2006) 79.

\end{thebibliography}
\end{document}